\documentclass[british]{article}
\usepackage[T1]{fontenc}
\usepackage[latin9]{inputenc}
\usepackage{geometry}
\usepackage{color}
\usepackage{babel}
\usepackage{float}
\usepackage{textcomp}
\usepackage{url}
\usepackage{amsmath}
\usepackage{amssymb}
\usepackage{graphicx}
\usepackage[unicode=true,
 bookmarks=true,bookmarksnumbered=false,bookmarksopen=false,
 breaklinks=true,pdfborder={0 0 0},pdfborderstyle={},backref=false,colorlinks=true]
 {hyperref}
\hypersetup{pdftitle={KinForm: Kinetics Informed Feature Optimised Representation Models for Enzyme k_{cat} and K_{M} Prediction},
 pdfauthor={Saleh Alwer},
 colorlinks=true,citecolor=magenta,linkcolor=blue}

\makeatletter
\newenvironment{lyxcode}
	{\par\begin{list}{}{
		\setlength{\rightmargin}{\leftmargin}
		\setlength{\listparindent}{0pt}
		\raggedright
		\setlength{\itemsep}{0pt}
		\setlength{\parsep}{0pt}
		\normalfont\ttfamily}%
	 \item[]}
	{\end{list}}

\usepackage[labelfont=bf]{caption}
\usepackage{geometry}
\usepackage{lmodern}
\usepackage{setspace}
\usepackage{bbm}

\author{
    Saleh Alwer\textsuperscript{1,2} \and
    Ronan Fleming\textsuperscript{1,2}\thanks{Corresponding author: \texttt{ronan.mt.fleming@gmail.com}}
}

\date{
    \textsuperscript{1}Digital Metabolic Twin Center, University of Galway, Galway, Ireland\\
    \textsuperscript{2}School of Medicine, University of Galway, Galway, Ireland
}

\makeatother

\usepackage[numbers]{natbib}
\begin{document}
\title{KinForm: Kinetics-Informed Feature Optimised Representation Models
for Enzyme $k_{cat}$ and $K_{M}$ Prediction}
\maketitle
\begin{abstract}
Kinetic parameters such as the turnover number ($k_{cat}$) and Michaelis
constant ($K_{\mathrm{M}}$) are essential for modelling enzymatic
activity but experimental data remains limited in scale and diversity.
Previous methods for predicting enzyme kinetics typically use mean-pooled
residue embeddings from a single protein language model to represent
the protein. We present KinForm, a machine learning framework designed
to improve predictive accuracy and generalisation for kinetic parameters
by optimising protein feature representations. KinForm combines several
residue-level embeddings (Evolutionary Scale Modeling Cambrian, Evolutionary
Scale Modeling 2, and ProtT5-XL-UniRef50), taken from empirically
selected intermediate transformer layers and applies weighted pooling
based on per-residue binding-site probability. To counter the resulting
high dimensionality, we apply dimensionality reduction using principal--component
analysis (PCA) on concatenated protein features, and rebalance the
training data via a similarity-based oversampling strategy. KinForm
outperforms baseline methods on two benchmark datasets. Improvements
are most pronounced in low sequence similarity bins. We observe improvements
from binding-site probability pooling, intermediate-layer selection,
PCA, and oversampling of low-identity proteins. We also find that
removing sequence overlap between folds provides a more realistic
evaluation of generalisation and should be the standard over random
splitting when benchmarking kinetic prediction models.
\end{abstract}
\pagebreak{}

\section{Introduction}

\paragraph*{Background and Motivation}

Kinetic parameters such as the turnover number ($k_{cat}$) and the
Michaelis constant ($K_{\mathrm{M}}$) are fundamental to understanding
enzyme efficiency, substrate affinity, and the dynamics of biochemical
pathways. Despite their importance, experimentally determined values
are available for only a small fraction of known enzymes, and these
measurements are concentrated on a few well-studied protein families.
This limited and biased coverage constrains both basic research and
applied efforts in metabolic engineering. To address this, machine
learning (ML) models have emerged as a useful tool for predicting
kinetic parameters directly from protein sequence and molecular features.
These models rely on the featurisation of proteins and substrates,
making the design and optimisation of input representations a critical
determinant of predictive accuracy.

\paragraph*{Pre-trained Language Models for Protein Representation }

Recent efforts have leveraged pre-trained protein language models
(PLLMs), such as Evolutionary Scale Modelling 2 (ESM2) \cite{lin_evolutionary-scale_2023lin_evolutionary-scale_2023}
and ProtT5 \cite{pokharel_improving_2022pokharel_improving_2022},
to generate fixed-length protein embeddings from amino acid sequences.
These models, trained on millions of unlabelled sequences, provide
rich representations that encode structural and functional information.
PLLMs are trained in a self-supervised manner on large-scale unlabelled
data; they do not require kinetic annotations to learn biologically
meaningful features. This makes them especially valuable for downstream
tasks like kinetic parameter prediction, where labelled data is limited.
These models offer residue-level embeddings that capture evolutionary
and biochemical features learned from millions of unlabelled sequences.
Existing kinetics predictors such as DLKcat \cite{li_deep_2022li_deep_2022a},
TurNup \cite{kroll_turnover_2023kroll_turnover_2023a}, and UniKP
\cite{yu_unikp_2023yu_unikp_2023} reduce these embeddings to a single
per-protein vector by mean-pooling the last transformer layer before
feeding the result into a regression model.

\paragraph*{Outstanding Challenges }

The current convention has three limitations. First, catalytic activity
is governed by a small subset of residues so uniform pooling may dilute
relevant signals. Second, the last transformer layer is not invariably
the most informative for downstream prediction. Third, many studies
use random train--test splits that retain identical proteins on both
sides of the partition, inflating apparent performance and obscuring
a model\textquoteright s ability to generalise to unseen sequences
\cite{kroll_turnover_2023kroll_turnover_2023a,kroll_dlkcat_2024kroll_dlkcat_2024}.

\paragraph*{Contributions of This Work }

We introduce KinForm, a framework that addresses the limitations above
while maintaining computational efficiency. Specifically, we (i) weight
residue embeddings by predicted binding-site probabilities to emphasise
catalytically relevant positions, (ii) select task-specific intermediate
layers from multiple PLLMs rather than defaulting to the last transformer
layer, (iii) control overfitting through principal component analysis
(PCA) and similarity-based oversampling of low-identity sequences,
and (iv) adopt Sequence-Exclusive Cross Validation (SE-CV), which
eliminates sequence overlap between training and test sets, for more
accurate evaluation of generalisation. Together, these elements improve
prediction accuracy, particularly for low-similarity proteins, without
requiring structural input or specialised hardware.

\section{Methods}

\subsection{Overview}

We summarise the KinForm architecture in Figure \ref{fig:KINEFORM-Framework.}.
The protein sequence is embedded using three protein language models:
ESMC \cite{esm2024cambrian}, ESM-2 \cite{lin_evolutionary-scale_2023lin_evolutionary-scale_2023},
and ProtT5-XL \cite{pokharel_improving_2022pokharel_improving_2022},
each producing residue-level representations. We extract embeddings
from model-specific layers that yield strong performance. We then
aggregate the residue-level embeddings from each language model $j$
in two ways: (1) a global mean across all residues, yielding $\mathbf{m}_{g}^{(j)}\in\mathbb{R}^{H_{j}}$,
and (2) a binding-site--weighted mean, yielding $\mathbf{m}_{b}^{(j)}\in\mathbb{R}^{H_{j}}$,
where $H_{j}$ denotes the hidden dimension of model $j$. The weights
used to compute $\mathbf{m}_{b}^{(j)}$ are the normalised per-residue
probabilities of binding-site involvement, as predicted by Pseq2Sites
\cite{seo_pseq2sites_2024seo_pseq2sites_2024}. This results in six
vectors of dimension $H_{j}$, with one global mean and one binding-site--weighted
vector computed for each model $j$ (Section \ref{subsec:Protein-Representations}).

For the generalisation model,\emph{ }KinForm-L\emph{ }(the ``L''
stands for low similarity, indicating generalisation beyond the training
distribution)\emph{, }each of the six vectors is independently scaled
using the median and interquartile range to ensure robustness to outliers.
The scaled global and binding-weighted vectors are separately concatenated
to form $\mathbf{m}_{g},\mathbf{m}_{b}\in\mathbb{R}^{3456}$. Each
of the concatenated vectors is then standardised using $z$-score
normalisation, ensuring zero mean and unit variance and then reduced
to $300$ dimensions via PCA fit on the training set. The resulting
protein representation is a $600$-dimensional vector, obtained by
concatenating the reduced global and binding components (Section \ref{subsec:Dimensionality-Reduction-via}).
The protein vector is then combined with a molecular embedding derived
from the substrate SMILES (Section \ref{subsec:SMILES-Representation})
and used as input to an ensemble-based regression method that constructs
multiple unpruned decision trees with randomised splits, known as
Extra Trees \cite{geurts_extremely_2006geurts_extremely_2006}, for
predicting $k_{cat}$ or $K_{\mathrm{M}}$. To mitigate the dominance
of highly similar sequence clusters in the training data, we apply
a clustering-based oversampling strategy (Section \ref{subsec:Oversampling-Based-on}).

For the variant targeting in-distribution inputs, KinForm-H (where
\textquoteleft H\textquoteright{} stands for high similarity, reflecting
better performance on training-like or in-distribution data), the
six embedding vectors are concatenated directly, without scaling or
PCA, yielding a 6912-dimensional protein representation. This full-resolution
representation is likewise concatenated with the molecular embedding
before being passed to the same regression model. The SMILES embeddings
are derived from the same SMILES Transformer \cite{honda_smiles_2019honda_smiles_2019}
used by UniKP. Extra Trees is used for regression because it is the
top-performing model in a benchmark of 16 machine learning algorithms
reported by UniKP. The configurations of KinForm--H and KinForm-L
are described in Section \ref{subsec:Model-Configurations}.
\begin{lyxcode}
\begin{center}
\begin{figure}[H]
\centering{}\includegraphics[width=1\textwidth]{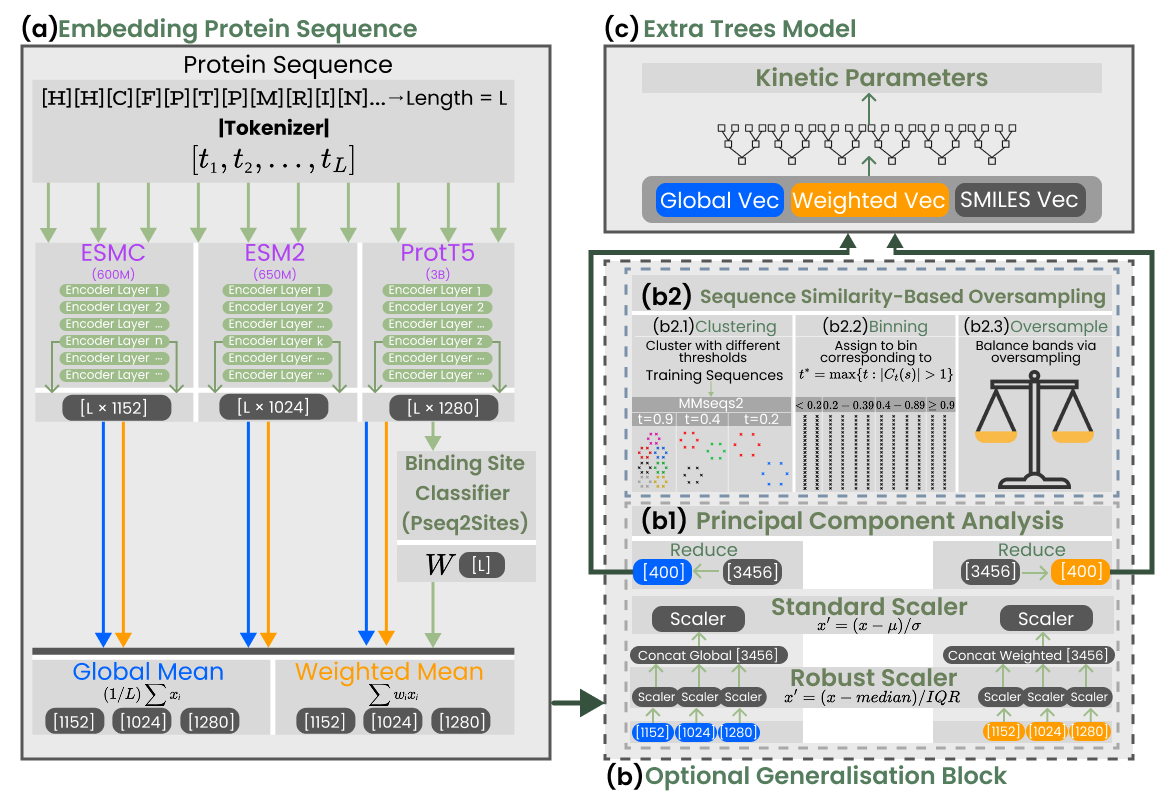}\caption{\label{fig:KINEFORM-Framework.}\textbf{Overview of the KinForm Framework.
(a)} The input protein sequence is tokenised and processed by three
pre-trained language models: ESMC, ESM-2, and ProtT5, with approximately
600M, 650M, and 3B parameters, respectively. Residue-level embeddings
are extracted from empirically selected intermediate layers of each
model, chosen separately for each task ($K_{\mathrm{M}}$ or $k_{cat}$).
Binding site probabilities are predicted using Pseq2Sites. For each
model, residue embeddings are aggregated via two approaches: (i) a
simple mean to obtain a global vector $\mathbf{m}_{g_{i}}\in\mathbb{R}^{H}$,
and (ii) a binding-weighted mean using the predicted site probabilities
to obtain $\mathbf{m}_{b_{i}}\in\mathbb{R}^{H}$, where $H$ is the
hidden dimension of model $i$.\textbf{ (b1)} The six resulting vectors
($m_{b_{i}}$ and $m_{g_{i}}$for each PLLM $i$) are independently
scaled. Global and binding-weighted vectors are separately concatenated
to form $\mathbf{m}_{g},\mathbf{m}_{b}\in\mathbb{R}^{3456}$, standardised
via $z$-score normalisation, and reduced from $3456$ to $300$ dimensions
using PCA.\textbf{ (b2)} Oversampling based on sequence similarity
distribution.\textbf{ (b2.1) }Training sequences are clustered at
three identity thresholds. \textbf{(b2.2) }Each sequence is assigned
to the highest threshold $t$, where its cluster size exceeds $1$.\textbf{
(b2.3) }Lower bands are over-sampled to match the size of the $\ge0.9$\textbf{
}band. \textbf{(c)} The resulting protein representation is concatenated
with a SMILES-derived embedding of the small molecule, used as input
to an Extra Trees regressor to predict $k_{cat}$ or $K_{\mathrm{M}}$.}
\end{figure}
\par\end{center}
\end{lyxcode}

\subsection{Datasets}

We use two publicly available $k_{cat}$ datasets: DLKcat \cite{li_deep_2022li_deep_2022a}
and the Shen \cite{shen_eitlem-kinetics_2024shen_eitlem-kinetics_2024}
dataset, each compiled and released by their respective publications.
Both datasets are sourced from BRENDA \cite{chang_brenda_2021chang_brenda_2021a}
and SABIO-RK \cite{wittig_sabio-rk_2018wittig_sabio-rk_2018a}, containing
experimentally measured turnover numbers across diverse enzyme--substrate
pairs. We observe that DLKcat and Shen share similar distributions
of $k_{cat}$ values and wild type versus mutant proportions and almost
the entirety of the DLKcat dataset is contained within Shen (Supplementary
Figure \ref{fig:Dataset-comparison}). Despite this, we still use
the DLKcat dataset as a separate evaluation set because it was the
dataset used in the original UniKP paper, which serves as the main
baseline model in our comparisons. Both datasets exhibit comparable
sequence redundancy patterns, with similar distributions of sequence
identity--based cluster sizes and sequence overlap across clusters,
but the Shen dataset spans a significantly broader sequence space,
capturing roughly twice the effective diversity of DLKcat, while maintaining
comparable levels of redundancy (Supplementary Figure \ref{fig:Dataset-sim}).
After removing negative $k_{cat}$ values and invalid SMILES, the
DLKcat dataset contains $16,775$ $k_{cat}$ datapoints, and the Shen
dataset contains $35,001$ $k_{cat}$ datapoints. The $K_{\mathrm{M}}$
dataset in this work is the same one used by UniKP \cite{kroll_deep_2021kroll_deep_2021,yu_unikp_2023yu_unikp_2023}.
This dataset consists of $11,722$ samples. All values, in the $k_{cat}$
and $K_{\mathrm{M}}$ datasets, are $\log_{10}$ transformed. All
$k_{cat}$ values are reported in $s^{-1}$, and all $K_{\mathrm{M}}$
values are in millimolar ($mM$).

We use two cross validation (CV) strategies for evaluation. The first
is standard 5-fold CV, where the dataset is randomly partitioned into
five folds such that each datapoint appears in the test set exactly
once. The second is 5-fold Sequence-Exclusive Cross Validation (SE-CV),
where we enforce that no protein sequence in the training set appears
in the test set within any fold. This stricter validation scheme evaluates
the model's ability to generalise to unseen protein sequences. For
each experiment, a fixed random seed is used to ensure that all model
configurations are evaluated on the same data splits. Across different
experiments, we vary the seed to avoid biasing results toward a particular
split configuration.

\subsection{\label{subsec:Protein-Representations}Protein Representations}

We use three pretrained protein language models to extract residue-level
embeddings: ESMC (600M), ESM-2 (650M), and ProtT5-XL-UniRef50 (3B).
ESM-2 is accessed via the official ESM library provided by Facebook
AI \cite{lin_evolutionary-scale_2023lin_evolutionary-scale_2023},
which allows model initialisation and loading of pre-trained weights.
ESMC is accessed via a separate Python package provided by EvolutionaryScale
\cite{evolutionaryscale_2024}. ProtT5-XL is accessed through the
Hugging Face Transformers library \cite{wolf_huggingface_2020wolf_huggingface_2020}.
Each model takes as input a FASTA-formatted amino acid sequence. After
model-specific tokenisation and encoding, each sequence is passed
through the model's encoder stack to obtain a tensor of hidden states
$\mathbf{E}\in\mathbb{R^{N\times L\times H}}$, where $N$ is the
number of layers ($34$ for ESM-2, $24$ for ProtT5-XL, and $36$
for ESMC), $L$ is the sequence length, and $H$ is the hidden dimension
of the model ($1280$ for ESM-2, $1024$ for ProtT5-XL, and $1152$
for ESMC).

We then apply mean pooling across the sequence length dimension for
each layer resulting in a matrix $\mathbf{M}\in\mathbb{R}^{N\times H}$,
as defined in Equation \ref{eq:mean-pooling-layerwise}

\begin{equation}
\mathbf{M}:=\begin{bmatrix}\mathbf{m}_{1}\\
\mathbf{m}_{2}\\
\vdots\\
\mathbf{m}_{N}
\end{bmatrix},\qquad\mathbf{m}_{n}:=\frac{1}{L}\mathbf{1}^{\top}\mathbf{E}_{n},\label{eq:mean-pooling-layerwise}
\end{equation}

\noindent where $\mathbf{E}_{n}\in\text{\ensuremath{\mathbb{R}^{L\times H}}}$
is the matrix of residue embeddings at layer $n$, and $\mathbf{1}\in\mathbb{R}^{L}$
is a column vector of ones used to average across residues. These
$N$ vectors are used in the layer-wise experiments described in Section
\ref{subsec:Predicted-Binding-Site-Weights} to identify the most
informative layers for kinetic parameter prediction. 

To generate fixed-length protein representations from residue-level
embeddings of a single layer $n$, we use three pooling strategies:
\emph{global}, \emph{binding-site weighted}, and \emph{both}. In the
\emph{global} setting, we compute the mean of the residue-level embeddings
across the full sequence, yielding a fixed-length vector $\mathbf{m}_{g}\in\mathbb{R}^{H}$
(Equation \ref{eq:global-pooling})

\begin{equation}
\mathbf{m}_{g}:=\frac{1}{L}\mathbf{1}^{\top}\mathbf{E}_{n}.\label{eq:global-pooling}
\end{equation}

In the \emph{binding-site weighted} setting, we incorporate residue-level
binding-site probabilities predicted by Pseq2Sites. Given a vector
of binding-site probabilities, $\mathbf{w}\in[0,1]^{L}$, where each
entry $w_{i}$ represents the predicted probability that residue $i$
belongs to the binding site, we normalise so the values sum to $1$
across the protein sequence (Equation \ref{eq:normalize-binding})

\begin{equation}
\tilde{\mathbf{w}}:=\frac{\mathbf{w}}{\mathbf{1}^{\top}\mathbf{w}}.\label{eq:normalize-binding}
\end{equation}

\noindent The \emph{binding-site weighted} representation, $\mathbf{m}_{b}\in\mathbb{R}^{H}$
is then computed as a weighted sum over the residue embeddings using
the normalised weights (Equation \ref{eq:binding-pool})

\begin{equation}
\mathbf{m}_{b}:=\tilde{\mathbf{w}}^{\top}\mathbf{E}_{n}.\label{eq:binding-pool}
\end{equation}

In the \emph{both} setting, we concatenate the \emph{global} and \emph{binding-site--weighted}
vectors to form a combined representation $\mathbf{m}_{gb}\in\text{\ensuremath{\mathbb{R}^{2H}}}$(Equation
\ref{eq:concat_gb})

\begin{equation}
\mathbf{m}_{gb}:=[\mathbf{m}_{\text{g}}\,\|\,\mathbf{m}_{\text{b}}],\label{eq:concat_gb}
\end{equation}

\noindent where $\mathbf{m}_{\text{g}},\mathbf{m}_{\text{b}}\in\mathbb{R}^{H}$
are the \emph{global} and\emph{ binding-site--weighted} representations,
respectively, and $\|$ denotes vector concatenation. When using multiple
protein language models, we compute $\mathbf{m}_{g}$ and $\mathbf{m}_{b}$
separately for each model and concatenate them. In this case, $H$
is the sum of the hidden dimensions of the individual models.

For each protein language model and kinetic parameter, we selected
a single layer for extracting embeddings. To choose these layers,
we evaluated layer-wise performance using a predefined composite score
$S$ that balances central tendency and stability (Equation \ref{eq:optimal_layer})

\begin{equation}
S:=0.8\,\widetilde{R^{2}}+0.4\,\overline{R^{2}}-0.5\,\sigma_{R^{2}},\label{eq:optimal_layer}
\end{equation}

\noindent where $\widetilde{R^{2}}$ is the median, $\overline{R^{2}}$
the mean and $\sigma_{R^{2}}$ the standard deviation of $R^{2}$
scores across folds. The weights were not systematically tuned, but
chosen based on heuristics to prioritise layers with strong average
and median performance and low variance. Section \ref{subsec:Best-Performing-Representations-}
reports the full evaluation and selected layers.

\subsection{\label{subsec:Dimensionality-Reduction-via} Dimensionality Reduction}

The feature dimension of a protein using the three PLLMs and global
and binding representations is $6912$. This results from concatenating
six individual embedding blocks, each of which is derived from the
same protein sequence, introducing redundancy that may cause ensemble
tree methods to overfit. To reduce the dimensionality of protein representations
while preserving their information content, we apply PCA after a two
stage scaling process. 

Protein representations consist of six individual blocks; one global
and one binding-weighted vector from each of the three language models
(ESMC, ESM2, ProtT5). We denote each block as defined in Equation
\ref{eq:prot_blocks}

\begin{equation}
\mathbf{X}_{m}^{t}\in\mathbb{R}^{n\times d_{t}},\label{eq:prot_blocks}
\end{equation}

\noindent where $m\in\{g,b\}$ refers to global or binding vectors
and $t\in\{ESMC,ESM2,T5\}$, is the source model, $n$ is the number
of datapoints and $d_{t}$ is the hidden dimension of the model $t$.
Each block is first independently scaled using robust statistics,
defined in Equation \ref{eq:robust_scale}, computed from the training
set

\begin{equation}
\mathbf{X}_{m}^{t\prime}:=\frac{\mathbf{X}_{m}^{t}-\text{median}(\mathbf{X}_{m,\text{train}}^{t})}{\text{IQR}(\mathbf{X}_{m,\text{train}}^{t})},\label{eq:robust_scale}
\end{equation}

\noindent where $median(\cdot)$ and $IQR(\cdot)$ denote element-wise
median and interquartile range, respectively. This scaling is resilient
to outliers and ensures that blocks with high variance dominate the
PCA transformation. 

The three global vectors are concatenated to form $\mathbf{X}_{g}\in\mathbb{R}^{n\times3456}$,
and the three binding-weighted vectors are similarly concatenated
to form $\mathbf{X}_{b}\in\mathbb{R}^{n\times3456}$ . Each matrix
is then standardised via $z$-score normalisation as defined in Equation
\ref{eq:standard_scale}

\begin{equation}
\mathbf{X}_{*}'':=\frac{\mathbf{X}_{*}-\mu(\mathbf{X}_{*,\text{train}})}{\sigma(\mathbf{X}_{*,\text{train}})}\quad\text{for }*\in\{g,b\},\label{eq:standard_scale}
\end{equation}

\noindent where $\mu$ and $\sigma$ are the mean and standard deviation
computed from the training set.

PCA is applied separately to the standardised global and binding matrices,
reducing each to $k$ dimensions. We denote the reduced representation
of each matrix as $\mathbf{Z}_{*}=\mathrm{PCA}_{*}(\mathbf{X}_{*}'')\in\mathbb{R}^{n\times k}\quad\text{for }*\in\{g,b\}$.
The final protein representation is the concatenation of the reduced
global and binding components defined in Equation \ref{eq:concat_Z}.
All transformations (scaling and PCA) are fit only on the training
set and reused to transform the test set in the same sequence. Specifically,
for each block, the test data is first robustly scaled using the training
medians and IQRs, then concatenated, standardised using the training
means and variances, and finally projected into the reduced PCA space
using the components fit on the training data

\begin{equation}
\mathbf{Z}:=[\mathbf{Z}_{g}\,|\,\mathbf{Z}_{b}]\in\mathbb{R}^{n\times2k}.\label{eq:concat_Z}
\end{equation}

\subsection{\label{subsec:Oversampling-Based-on}Oversampling Based on Sequence
Similarity}

To counteract training biases introduced by over-represented, highly
similar protein sequences in the training set and to improve generalisation,
we implement a clustering-based oversampling strategy. This procedure
is applied exclusively to the training split within each cross validation
fold.

We begin by identifying unique protein sequences in the training data
and clustering them using MMseqs2 at three identity thresholds: $0.90$
, $0.40$ , and $0.20$. Each unique sequence is assigned to the highest
identity threshold $t\in\{0.90,0.40,0.20\}$ for which its cluster
contains more than one member. Formally, for a given sequence $s$,
its assigned threshold is defined in Equation \ref{eq:assign_thres}

\begin{equation}
t^{*}:=\max\{t:|C_{t}(s)|>1\},\label{eq:assign_thres}
\end{equation}

\noindent where $C_{t}(s)$ denotes the denotes the MMseqs2 cluster
of $s$ at threshold $t$. If a sequence is a singleton in all clusters,
it is assigned to the $<0.20$ bin. Following this assignment, all
training sequences are grouped by their corresponding similarity bands:
$<0.2$, $0.20-0.39$, $0.40-0.89$ and $\ge0.90$. We then oversample
the lower bands to match the number of samples in the $\ge0.90$ bin.
Oversampling is performed by random duplication with replacement of
the original training datapoints within each band.

\subsection{\label{subsec:SMILES-Representation}Molecule Representation}

For substrate representation, we use embeddings from the SMILES Transformer
\cite{honda_smiles_2019honda_smiles_2019} due to its robust performance
across tasks (Supplementary Figure \ref{fig:Comparison-of-SMILES}).
This approach is also used by UniKP. The input SMILES is tokenised
and passed through the model's encoder stack. From both the last encoder
layer $\ell$ and the penultimate layer $\ell-1$, we extract token-level
embeddings $\mathbf{H}^{(\ell)}\in\mathbb{R}^{T\times d}$ and $\mathbf{H}^{(\ell-1)}\in\mathbb{R}^{T\times d}$,
where $T$ is the number of SMILES tokens and $d=256$ is the hidden
dimension of the transformer. We then apply mean and max pooling over
the token dimension for each layer, denoted as $\overline{\mathbf{H}}^{(\ell)}$
and $\hat{\mathbf{H}}^{(\ell)}$, respectively. The resulting four
pooled vectors are concatenated to form the final SMILES embedding
$\mathbf{m}_{SMILES}\in\mathbb{R}^{1024}$ (Equation \ref{eq:smiles_transformer})

\begin{equation}
\mathbf{m}_{\text{SMILES}}:=\left[\overline{\mathbf{H}}^{(\ell-1)}\,\|\,\hat{\mathbf{H}}^{(\ell-1)}\,\|\,\overline{\mathbf{H}}^{(\ell)}\,\|\,\hat{\mathbf{H}}^{(\ell)}\right]\in\mathbb{R}^{1024}.\label{eq:smiles_transformer}
\end{equation}

\noindent We also evaluated several alternative substrate representations,
described below.

\paragraph{Uni-Mol2}

Molecules are first converted to 3D conformers and then passed through
the pretrained Uni-Mol2 model \cite{ji_exploring_2024ji_exploring_2024}.
From the last encoder layer, we extract the CLS token embedding $\mathbf{h}_{CLS}\in\mathbb{R}^{1536}$
and the mean of atom-level embeddings $mean_{atoms}\in\mathbb{R}^{1536}$.
These are concatenated to produce the final molecular embedding $\mathbf{m}_{\text{UniMol2}}\in\mathbb{R}^{3072}$.

\paragraph{Functional group-aware embeddings (FARM)}

Functional group--aware molecular embeddings are computed using the
pre-trained FARM model \cite{nguyen_farm_2024nguyen_farm_2024}. As
with Uni-Mol2, we concatenate the CLS token embedding and the mean
of last-layer atom embeddings to yield $\mathbf{m}_{\text{FARM}}\in\mathbb{R}^{1536}$.

\paragraph{MolFormer }

The MolFormer model \cite{ross_large-scale_2022ross_large-scale_2022}
produces a fixed-length embedding by applying mean pooling over the
atom-level embeddings from the last encoder layer, yielding $\mathbf{m}_{\text{MolFormer}}\in\mathbb{R}^{768}$.

\paragraph{Fingerprint Based Representations}

We also evaluate classical molecular fingerprints, which are: 1) Morgan
fingerprint (MFP) \cite{morgan_generation_1965morgan_generation_1965},
2) Topological torsion (TT) fingerprint \cite{nilakantan_topological_1987nilakantan_topological_1987b},
3) MACCS keys \cite{durant_reoptimization_2002durant_reoptimization_2002},
4) Avalon fingerprint \cite{gedeck_qsar_2006gedeck_qsar_2006} and
5) MinHash fingerprint \cite{probst_probabilistic_2018probst_probabilistic_2018}.
Each method produces a fixed-length binary or integer vector representation
of the molecule. The dimension of the vector representing a molecule
produced by each method are: MFP ($2048$), TT ($2048$), MinHash
($2048$), MACCS ($167$), AtomPair ($2048$), and Avalon ($512$).
We implement a pipeline for encoding SMILES into these fingerprints
using Scikit-mol \cite{bjerrum_scikit-mol_2023} which we use to standardise
the inputs and convert to descriptors using RDKit \cite{landrum_rdkit_2025landrum_rdkit_2025}. 

To assess the effect of SMILES representations on predictive performance,
we evaluate the descriptors mentioned above. Each is used to generate
substrate representations, and an Extra Trees regressor is trained
using the same protein featurisation. Performance is evaluated using
both 5-fold CV and SE-CV. The SMILES Transformer yields the most consistent
performance across both metrics and tasks ($K_{m}$ and $k_{cat}$
) and is therefore used in all main experiments (Supplementary Figure
\ref{fig:Comparison-of-SMILES}).

\subsection{\label{subsec:Model-Configurations}Model Configurations}

All models used in this study follow a common structure: a fixed-length
protein representation is concatenated with a molecular embedding
derived from the substrate SMILES, and the resulting vector is passed
to an Extra Trees regressor to predict either $k_{cat}$ or $K_{\mathrm{M}}$.
All Extra Trees models were implemented using the Scikit-learn library
\cite{pedregosa_scikit-learn_2011pedregosa_scikit-learn_2011} and
trained using default hyperparameters. Below, we detail the configurations
of the models evaluated in this paper.

\subsubsection*{KinForm-H }

Residue-level embeddings are extracted from three pre-trained protein
language models (ESMC, ESM-2, and ProtT5-XL) using the optimal intermediate
layers identified in Section \ref{subsec:Best-Performing-Representations-}.
For each model, we compute both a global vector ($\mathbf{m}_{g}$)
and a binding-site weighted vector ($\mathbf{m}_{b}$) as defined
in Equations \ref{eq:global-pooling} and \ref{eq:binding-pool},
respectively. These vectors are concatenated for each model, and the
combined protein representation $\mathbf{m}_{\text{prot}}\in\mathbb{R}^{6912}$
is formed by stacking the global and binding vectors across all three
models (as described in Section \ref{subsec:Protein-Representations}).

The substrate is represented using the SMILES Transformer embedding
defined in Equation \ref{eq:smiles_transformer}. The final input
vector to the Extra Trees model is defined in Equation \ref{eq:kinformH_input}.
This configuration is used for both $K_{\mathrm{M}}$ and $k_{cat}$
prediction tasks.

\begin{equation}
\mathbf{x}_{\text{input}}:=[\mathbf{m}_{\text{prot}}\,\|\,\mathbf{m}_{\text{SMILES}}]\in\mathbb{R}^{7936}.\label{eq:kinformH_input}
\end{equation}

\subsubsection*{KinForm-L}

This configuration also uses global and binding-site weighted vectors
from all three protein language models. However, after robust scaling
and $z$-score normalisation, the three global vectors are concatenated
and reduced to $300$ dimensions via PCA (Section \ref{subsec:Dimensionality-Reduction-via}).
The same procedure is applied to the three binding-weighted vectors,
resulting in protein representation $\mathbf{Z}$ as defined in Equation
\ref{eq:concat_Z}.

The substrate is represented using the SMILES Transformer embedding
defined in Equation \ref{eq:smiles_transformer}. The final input
vector to the Extra Trees model is defined in Equation \ref{eq:kinformL_input}.
This configuration is only used for $k_{cat}$ prediction.

\begin{equation}
\mathbf{x}_{\text{input}}:=[\mathbf{Z}\,\|\,\mathbf{m}_{\text{SMILES}}]\in\mathbb{R}^{1624}.\label{eq:kinformL_input}
\end{equation}

\subsubsection*{UniKP Baseline }

This baseline is implemented following the original UniKP paper and
its publicly available codebase \cite{yu_unikp_2023yu_unikp_2023}.
Each protein is represented using the last encoder layer of ProtT5-XL,
aggregated via global mean pooling, resulting in $\mathbf{m}_{T5}\in\mathbb{R}^{1024}$.
The SMILES Transformer embedding $\mathbf{m}_{\text{SMILES}}\in\mathbb{R}^{1024}$
(defined in Equation \ref{eq:smiles_transformer}) is then concatenated
with the protein embedding, giving the final input vector defined
in \ref{eq:unikp_input}.

\begin{equation}
\mathbf{x}_{\text{input}}:=[\mathbf{m}_{\text{ProtT5}}\,\|\,\mathbf{m}_{\text{SMILES}}]\in\mathbb{R}^{2048}.\label{eq:unikp_input}
\end{equation}

\subsection{\label{subsec:Evaluation-Metrics}Evaluation Metrics}

To assess predictive performance, we compute the coefficient of determination,
defined as $R^{2}=1-\frac{\sum_{i=1}^{n}(y_{i}-\hat{y}_{i})^{2}}{\sum_{i=1}^{n}(y_{i}-\bar{y})^{2}}$
and the root mean squared error (RMSE), given by $\mathrm{RMSE}=\sqrt{\frac{1}{n}\sum_{i=1}^{n}(y_{i}-\hat{y}_{i})^{2}}$
where $y_{i}$ is the true log-transformed kinetic value for sample
$i$, $\hat{y}_{i}$ is the corresponding predicted value, $\bar{y}$
is the mean of the true values, and $n$ is the total number of samples.

In addition to computing overall $R^{2}$ and $RMSE$ for each fold,
we also evaluate performance stratified by sequence similarity between
test and training proteins. To do this, we cluster all protein sequences
(train + test) using MMseqs2 at identity thresholds of $20\%,50\%,70\%$
and $90\%$. For each test protein, we assign it to the highest threshold
at which it shares a cluster with at least one training protein. Sequences
that do not cluster with any training protein at $20\%$ or higher
are placed in the $<20\%$ bin. We then evaluate each similarity bin
separately, enabling a more granular analysis of generalisation across
similarity regimes. Datapoints with $100\%$ similarity (identical
sequences) are removed from the $>90\%$ bin.

\section{Results}

\subsection{\label{subsec:Best-Performing-Representations-}Best-Performing Representations
Are Found in Intermediate Transformer Layers }

To identify which layers of the pre-trained protein language models
are most informative for predicting enzyme kinetics, we evaluated
representations from each encoder layer of ESMC, ProtT5-XL, and ESM-2.
For each layer, residue-level embeddings were mean-pooled to form
a global protein vector, which was then concatenated with a fixed
SMILES representation of the substrate. The resulting feature vector
was used as input to an Extra Trees model trained separately for each
prediction task: $K_{\mathrm{M}}$ and $k_{cat}$. We ran 5-fold SE-CV,
where no protein sequence is shared between the training and test
sets within a fold and we used the DLKcat dataset for these experiments.

\begin{figure}[H]
\centering{}\includegraphics[width=1\textwidth]{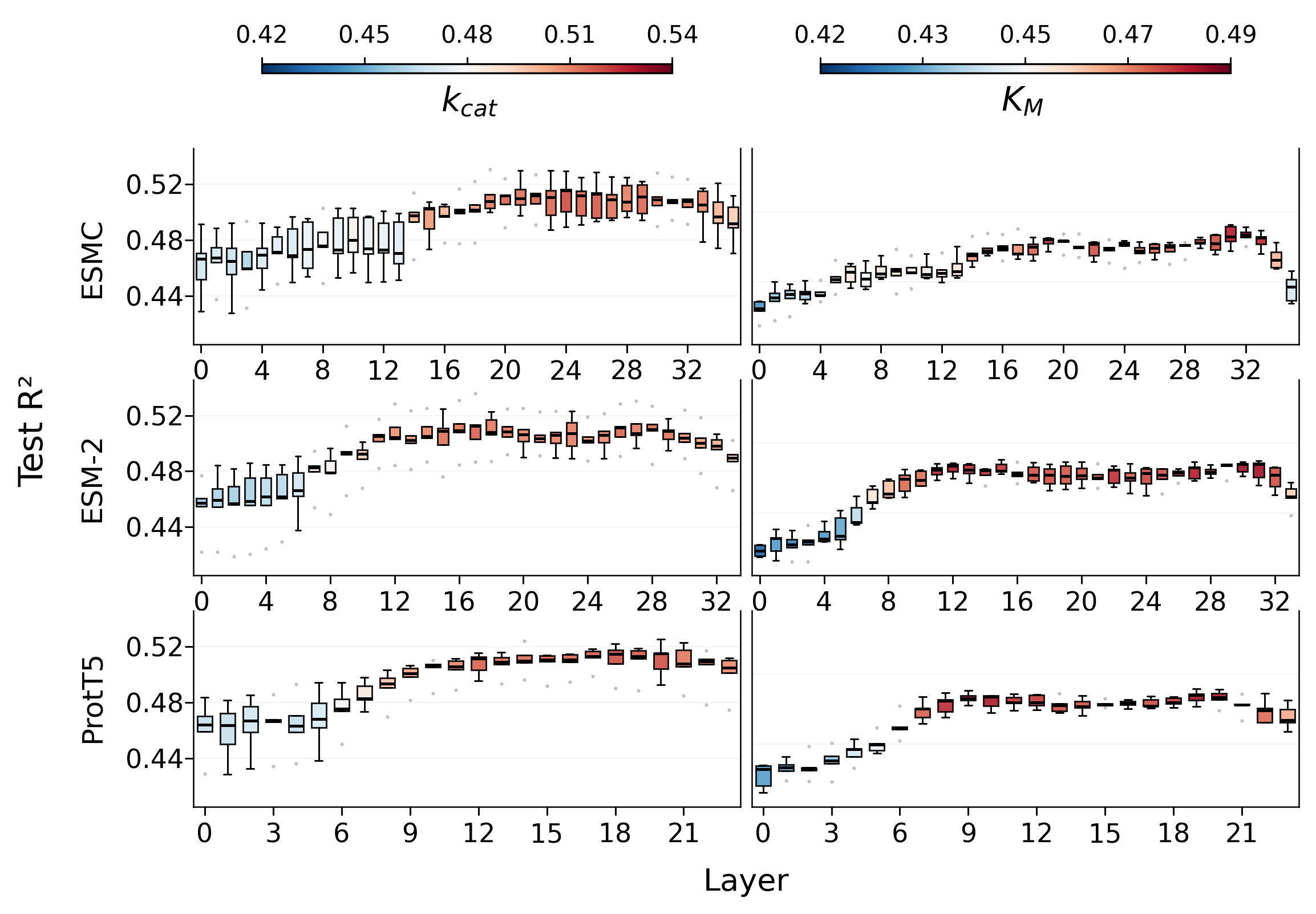}\caption{\label{fig:Layer-wise-performance-of}\textbf{Performance peaks at
intermediate layers rather than the last layer, and the optimal layer
varies across models and tasks. }The distribution of $R^{2}$ scores
across 5-fold SE-CV splits of Extra Trees models predicting $K_{m}$
and $k_{cat}$ using global pooled residue-level embeddings extracted
from each encoder layer of the three pre-trained models ESMC (36 layers),
ProtT5-XL (24 layers), and ESM2 (34 layers). In total, 940 Extra Trees
models were trained (one for each layer, task, and fold combination).
Colour bars represent the range of $R^{2}$ values for each task.
Grey dots represent outliers lying beyond $1.5\times IQR$. }
\end{figure}

The results, seen in Figure \ref{fig:Layer-wise-performance-of},
show that the highest predictive performance is often achieved by
intermediate layers rather than the last layer. Furthermore, the optimal
layer differs between the two tasks.

To identify a single layer per model for use in downstream experiments,
we computed a composite performance score incorporating median, mean,
and variance across folds (see Section \ref{subsec:Protein-Representations}).
Based on the criterion in Equation \ref{eq:optimal_layer}, we selected
the top-performing layer for each model--target pair. For ESM2, layer
$30$ for $K_{\mathrm{M}}$ and layer $26$ for $k_{cat}$ and for
ESMC, layer $32$ for $K_{\mathrm{M}}$ and layer $34$ for $k_{cat}$.
For ProtT5-XL layer $19$ is used for $K_{\mathrm{M}}$, while for
$k_{cat}$ we opted to retain the last encoder layer. Although a marginally
better score was observed in an intermediate layer, we chose the last
layer to align with UniKP's configuration, facilitating a fairer comparison
by isolating other architectural differences. These layers are used
throughout the paper unless otherwise specified.

\subsection{KinForm Improves Generalisation for $\mathbf{k_{cat}}$ Prediction}

We compared KinForm-L\emph{ }and KinForm-H to the UniKP baseline model
on two datasets: DLKcat and the Shen dataset. We report evaluation
using standard 5-fold CV and SE-CV. We reproduced the UniKP model
using the authors' published source code \cite{yu_unikp_2023yu_unikp_2023};
our results match those reported in the original paper. We exclude
the original DLKcat model from comparison and use UniKP as the primary
baseline, as it has been shown to substantially outperform DLKcat
on the same dataset. The configuration of all models evaluated here
is described in Section \ref{subsec:Model-Configurations}.

\begin{figure}[H]
\begin{centering}
\includegraphics[width=1\textwidth]{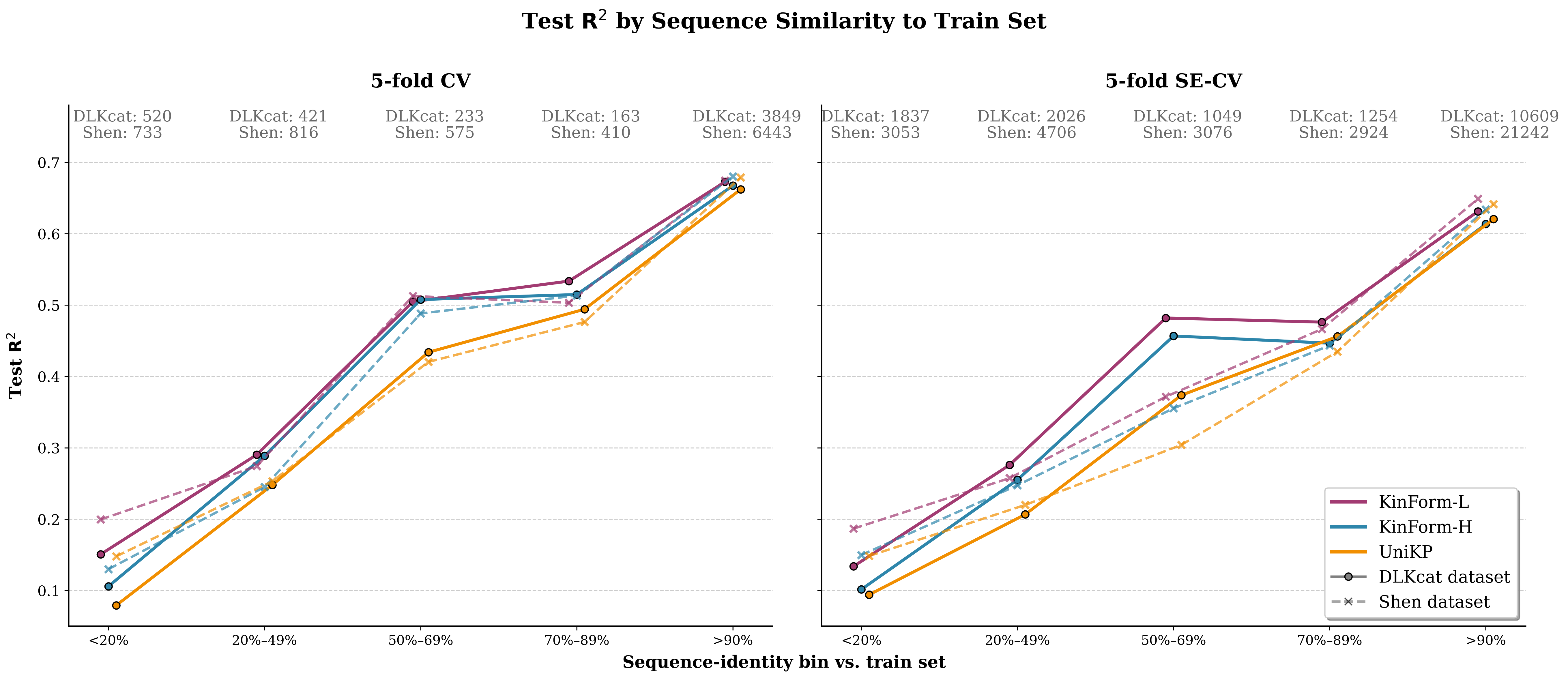}
\par\end{centering}
\centering{}\caption{\label{fig:seqsim-Vs-R2}\textbf{Models show strong performance on
high-identity sequences but degrade as similarity decreases. KinForm-L
consistently performs best on low-similarity inputs while at least
matching other models on high-similarity examples.} $R^{2}$ scores
across five sequence-identity bins under (left) standard 5-fold CV
and (right) SE-CV for each model. Solid lines correspond to the DLKcat
dataset; dashed lines to the Shen dataset. Numbers on the top indicate
the total number of test datapoints in the corresponding bin across
folds for each dataset. Repeated sequences ($100\%$ similarity) are
removed from the $>90\%$ bin in the CV plot. Lines are slightly horizontally
offset to improve readability and reduce overlap.}
\end{figure}

We further analysed test performance as a function of protein-sequence
identity to the training set (described in Section \ref{subsec:Evaluation-Metrics}).
As shown in Figure \ref{fig:seqsim-Vs-R2}, KinForm-L consistently
outperforms the other models in the lower similarity bins for both
datasets, providing additional evidence of improved generalisation
to out-of-distribution sequences. For instance, on the DLKcat dataset,
KinForm-L improves performance over UniKP in the $<20\%$ bin by $\approx100\%$
(from $0.077$ to $0.15$). We also observe large performance gains
in the $20$--$49\%$ and $50$--$69\%$ bins. The same trend is
observed on the Shen dataset and holds across both CV and SE-CV settings.

\begin{figure}[H]
\begin{centering}
\includegraphics[width=1\textwidth]{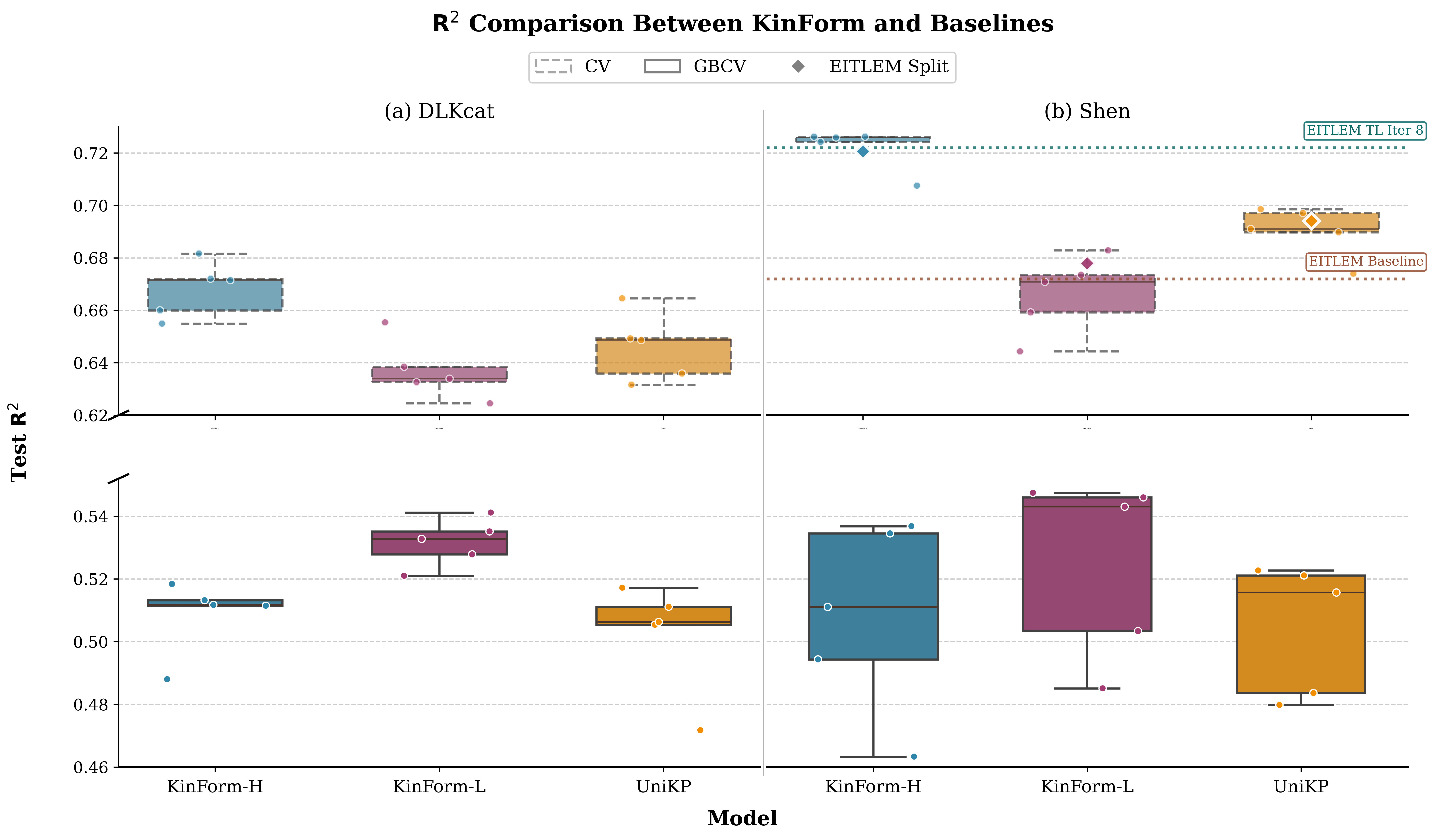}
\par\end{centering}
\centering{}\caption{\label{fig:unikp_comp_kcat}\textbf{ On both $\mathbf{k_{cat}}$ datasets,
KinForm-H achieves the highest performance under standard CV, while
KinForm-L performs best under SE-CV.} $R^{2}$ scores across five
folds for KinForm-L, KinForm-H, and UniKP under standard 5-fold CV
(dashed lines and lower opacity) and SE-CV (solid lines and full opacity).
Dots correspond to individual fold $R^{2}$.\textbf{ (a) }Results
on the DLKcat dataset and \textbf{(b) }Shen dataset. Dotted coloured
lines indicate the reported performance of the EITLEM model on its
original train/test split. \textquotedblleft Baseline\textquotedblright{}
is their reported performance after training on the $k_{cat}$ dataset
and \textquotedblleft TL Iter 8\textquotedblright{} is their reported
performance after eight iterations of transfer learning (final iteration).
Diamond markers represent the performance of each model when evaluated
on that same fixed reported split. EITLEM report $0.61$ test $R^{2}$
on a random 80/20 split of the DLKcat dataset. }
\end{figure}

Having established how performance varies with sequence similarity,
we now summarise overall predictive performance across the entire
test sets. As shown in Figure \ref{fig:unikp_comp_kcat}, KinForm-L\emph{
}achieves the highest performance under SE-CV on both datasets and
KinForm-H performs best under standard CV.

In the Shen dataset comparison (Figure \ref{fig:unikp_comp_kcat}(b)),
we additionally report the performance of the EITLEM model \cite{shen_eitlem-kinetics_2024shen_eitlem-kinetics_2024}
using the $80/20$ random split used in the original publication.
We did not re-train the EITLEM model; we use their reported results
and evaluate KinForm-L, KinForm-H, and UniKP on the same 80/20 published
split, enabling a direct comparison. Notably, KinForm-H\emph{ }achieves
an $R^{2}$ of $0.72$. This is comparable to the EITLEM model's reported
performance, which EITLEM\emph{ }achieved by pre-training a baseline
model and eight iterations of transfer learning. On the DLKcat dataset,
EITLEM\emph{ }reports an $R^{2}$ of 0.61 under a random $80/20$
split; all three models evaluated here, including UniKP, surpass that
baseline. On the Shen dataset, SE-CV results show smaller differences
in median performance between models and greater variability across
folds, whereas on the DLKcat dataset, model differences are larger
and fold-to-fold variability is lower.

\subsection{KinForm Outperforms UniKP Baseline for $\mathbf{K_{M}}$ Prediction}

We evaluate the KinForm model on $K_{\mathrm{M}}$ prediction using
the same 5-fold CV and SE-CV protocols as in the previous experiments,
and compare its performance to UniKP. For these experiments, we use
the full-resolution configuration of the model (previously referred
to as KinForm-H) and refer to it simply as KinForm. This is because
the dimensionality-reduced variant (KinForm-L) did not offer any performance
advantage on $K_{\mathrm{M}}$ under either cross validation setting.

\begin{figure}[H]
\begin{centering}
\includegraphics[width=1\textwidth]{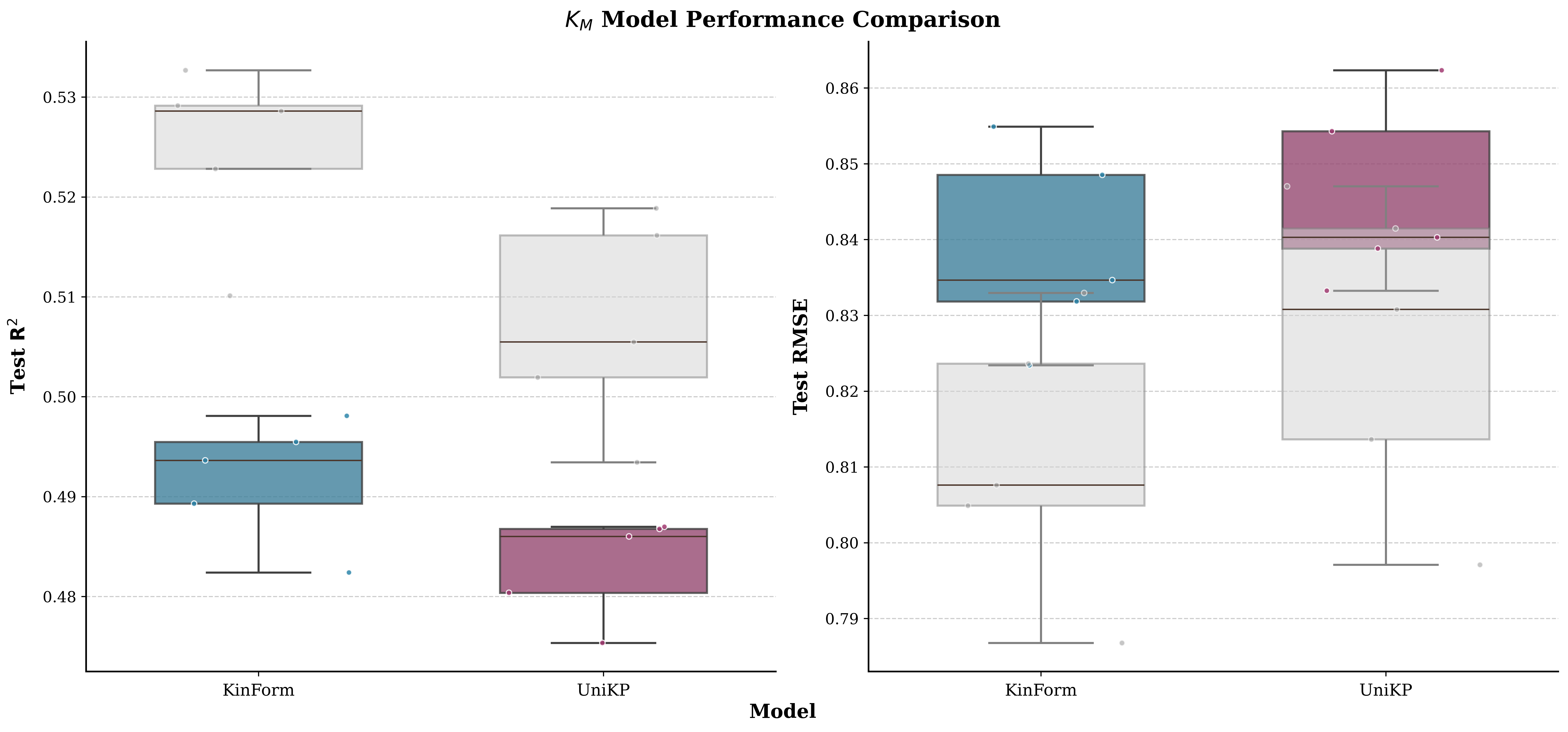}
\par\end{centering}
\centering{}\caption{\label{fig:km_comp}\textbf{Comparison of KinForm and UniKP on $K_{m}$
prediction.} $R^{2}$ (left) and $RMSE$ (right) values across five
folds under standard 5-fold CV (grey) and SE-CV (colour) of KinForm
and UniKP. Each dot represents the performance on a single fold.}
\end{figure}

KinForm outperforms UniKP in predicting $K_{\mathrm{M}}$ (Figure \ref{fig:km_comp}),
achieving higher $R^{2}$ and lower root mean squared error ($RMSE$
) under both CV and SE-CV settings. The performance gap is particularly
notable in the standard 5-fold CV setting. Compared to the $k_{cat}$
results, the difference between CV and SE-CV performance is smaller
here.

\subsection{Effect of Oversampling on Low-Similarity Performance}

\begin{figure}[H]
\begin{centering}
\includegraphics[width=1\textwidth]{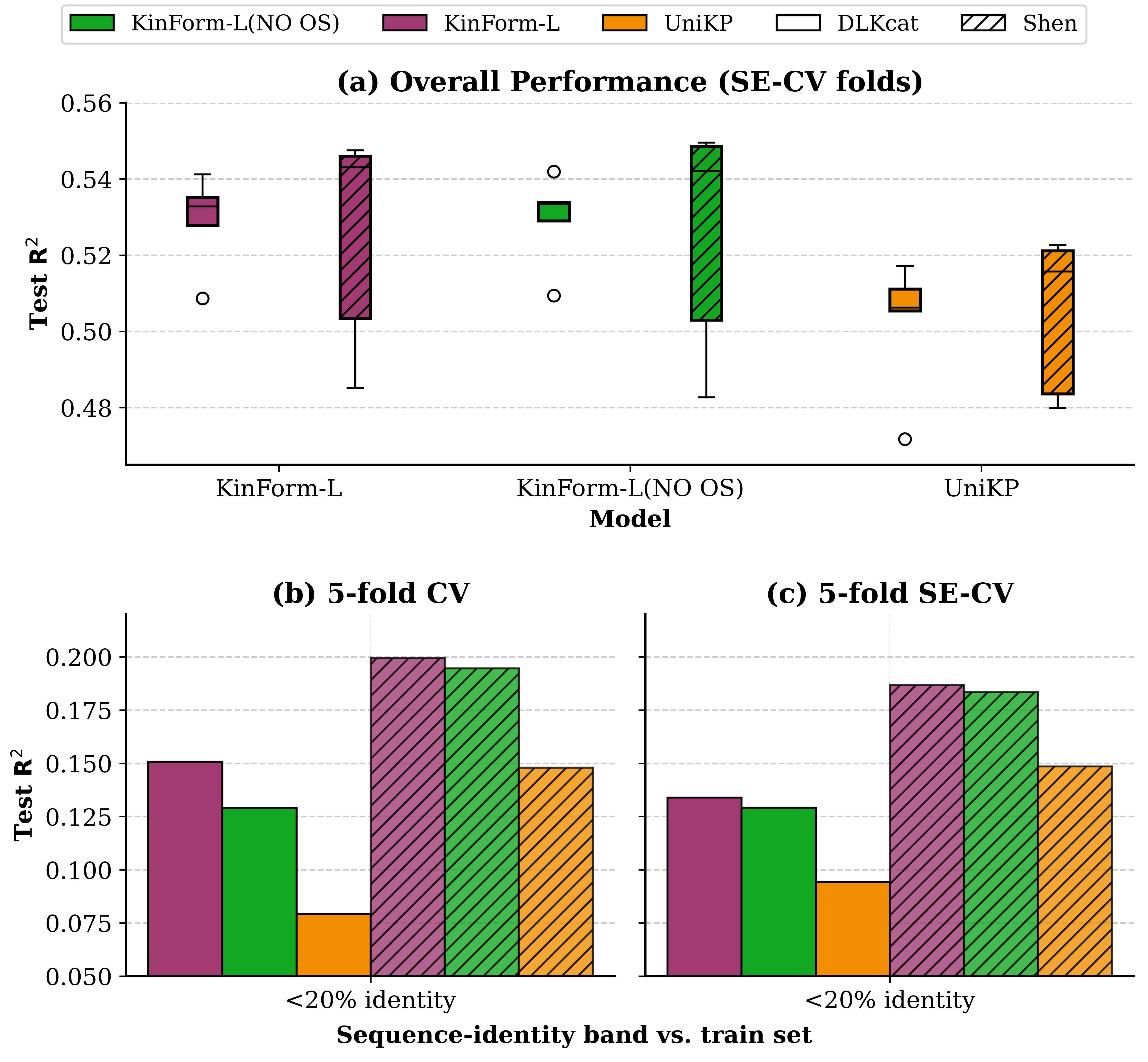}
\par\end{centering}
\centering{}\caption{\label{fig:OSvsNO(OS)}\textbf{Sequence similarity-based oversampling
improves generalisation. Oversampling has minimal impact on overall
performance, but improves generalisation to low-similarity sequences
under both CV and SE-CV. (a) }$R^{2}$ scores across five SE-CV folds
for KinForm-L with and without sequence similarity-based oversampling
(\emph{NO OS}), and the UniKP model on the DLKcat and Shen datasets.
\textbf{(b,c)} Comparison of $R^{2}$ scores in the $<30
$ sequence similarity bin under \textbf{(b)} standard 5-fold CV and
\textbf{(c)} 5-fold SE-CV. Bars are grouped by dataset and model;
colour denotes model configuration, and stripe fill indicates dataset.}
\end{figure}

To evaluate the specific effect of sequence similarity-based oversampling,
we ablated the similarity-aware oversampling step (No-OS) to isolate
its impact. As shown in Figure \ref{fig:OSvsNO(OS)}(a), overall performance
under 5-fold SE-CV remains similar between the two configurations
on both datasets, indicating that oversampling does not substantially
affect aggregate model accuracy. However, when performance is stratified
by sequence similarity, we observe a small but consistent improvement
in the $<20\%$ identity bin for the oversampled variant (Figure \ref{fig:OSvsNO(OS)}(b,c)).
These results suggest that oversampling improves prediction in low-similarity
regions while leaving high-similarity regions largely unaffected.

\subsection{\label{subsec:Predicted-Binding-Site-Weights}Performance Gains from
Binding-Site Aware Representations}

We assessed the impact of incorporating predicted binding-site information
by comparing three pooling strategies: global mean, binding-site--weighted
mean, and their concatenation. The definitions of these strategies
and their implementation are described in Section \ref{subsec:Protein-Representations}.
We conducted a grid search over seven combinations of pre-trained
protein models: ESMC, ESM-2, ProtT5-XL, and all pairwise and three-way
combinations among them. For each model combination and pooling strategy,
we trained an Extra Trees regressor ten times (5-fold CV and SE-CV),
each time concatenating the protein representation with the fixed
SMILES embedding of the substrate. We use the Shen dataset for these
experiments. The results on the DLKcat dataset are seen in Supplementary
Materials Figure \ref{fig:PCA_GS_EITLEM-1}.

\begin{figure}[H]
\centering{}\includegraphics[width=1\textwidth]{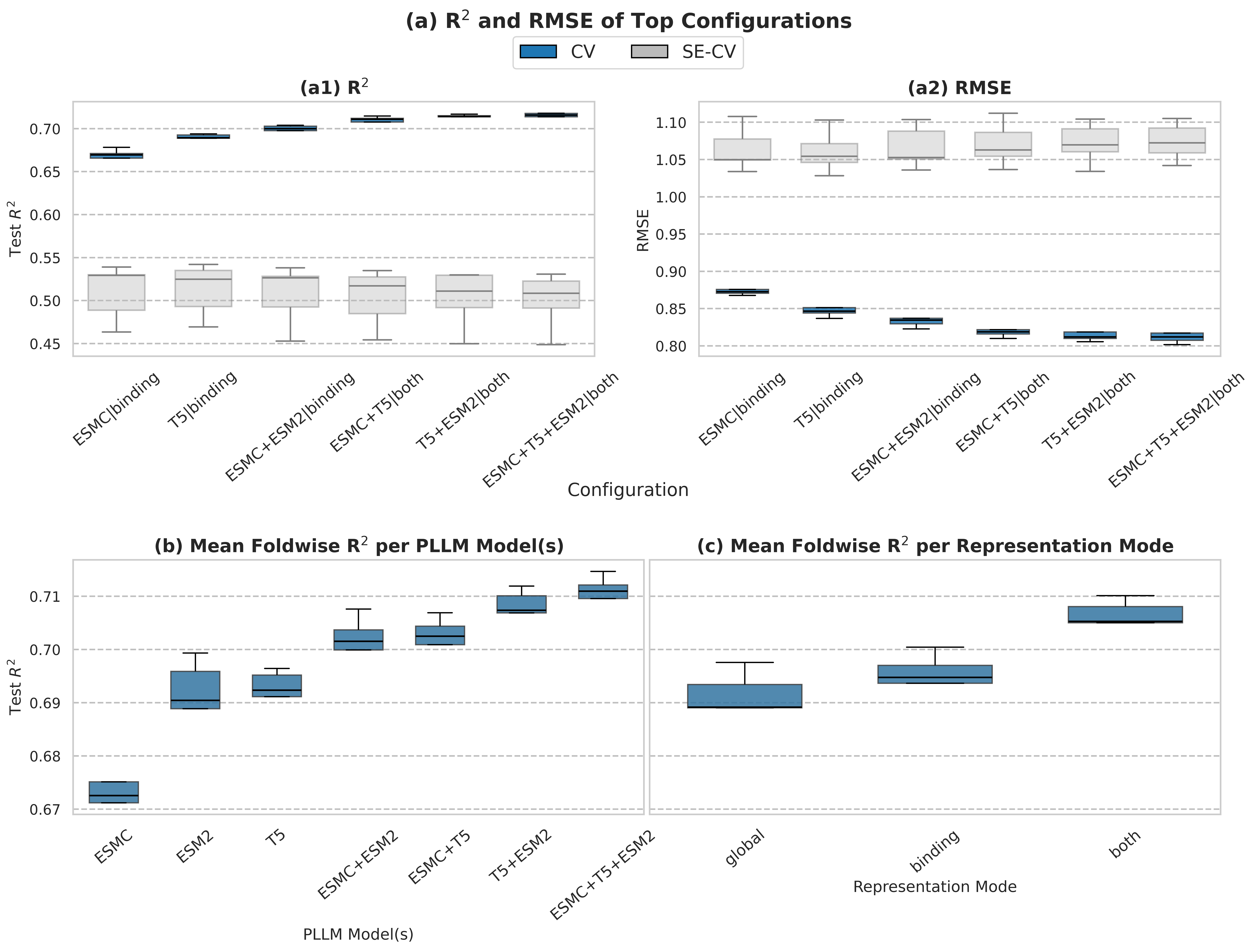}\caption{\label{fig:Effect-of-protein}\textbf{Top configurations improve standard
CV performance but not SE-CV. Fusing multiple PLLMs and binding-site--aware
representations outperforms simple mean pooling of one model. (a)}
$R^{2}$ \textbf{(a1) }and $RMSE$ \textbf{(a2)} scores across five
folds for the six top-performing protein representation configurations.
The x-axis indicates the protein language model(s) and representation
strategy used; global (mean pooling), binding (binding-site weighted
mean), or both (concatenation of global and binding). Blue boxes represent
standard CV results and grey boxes correspond to SE-CV. Shown are
the top 3 configurations by median $R^{2}$ according to 5-fold SE-CV
and the top 3 according to standard CV. Each configuration was trained
with an Extra Trees model and evaluated over 10 repeats (5-fold CV
and SE-CV).\textbf{ (b, c)} Distribution of average fold-wise $R^{2}$
(standard CV) scores for grouped sets of configurations. For each
group, we first compute the mean $R^{2}$ across folds for each configuration,
then aggregate these values across all configurations sharing the
same grouping. In \textbf{(b)}, configurations are grouped by protein
language model combination; in \textbf{(c)}, by protein representation
strategy. }
\end{figure}

Figure \ref{fig:Effect-of-protein} shows the best three configurations,
ranked separately by median $R^{2}$ under 5-fold CV and SE-CV. Two
main observations emerge from these results. First, all top configurations
use either the binding or both strategies, indicating that incorporating
binding-site weighting consistently improves performance compared
to using a standard global mean. Second, model combinations with higher
dimensional protein embeddings tend to improve CV performance, but
do not outperform single-model embeddings in SE-CV. 

\subsection{PCA Improves Out-of-Distribution $\mathbf{k_{cat}}$ Performance}

To evaluate the impact of PCA-based dimensionality reduction on generalisation,
we conducted experiments using two protein representation settings:
ESMC+ESM2+T5 and ESMC+T5. For each, we compared models trained using
the full high-dimensional protein vector against those using PCA-reduced
representations. PCA is applied after robust and standard scaling
(as shown in Figure \ref{fig:KINEFORM-Framework.} and expanded on
in Section \ref{subsec:Dimensionality-Reduction-via}), with component
counts ranging from 100 to 1750, selected to capture increasing levels
of cumulative variance. The cumulative explained variance captured
by PCA as a function of the number of components, computed separately
for global and binding representations, is shown in Supplementary
Figure \ref{fig:PCA-Var}.

Each representation, concatenated with the SMILES vector, was used
to fit an Extra Trees regressor to predict $k_{cat}$. We evaluated
performance using both 5-fold standard CV and SE-CV. We report results
on the DLKcat dataset for this experiments. The results on the Shen
dataset exhibit the same overall trends as observed here, though with
greater variability across folds and smaller margins between configurations
under 5-fold SE-CV (Supplementary Materials Figure \ref{fig:PCA_GS_EITLEM}). 

\begin{figure}[H]
\begin{centering}
\includegraphics[width=1\textwidth]{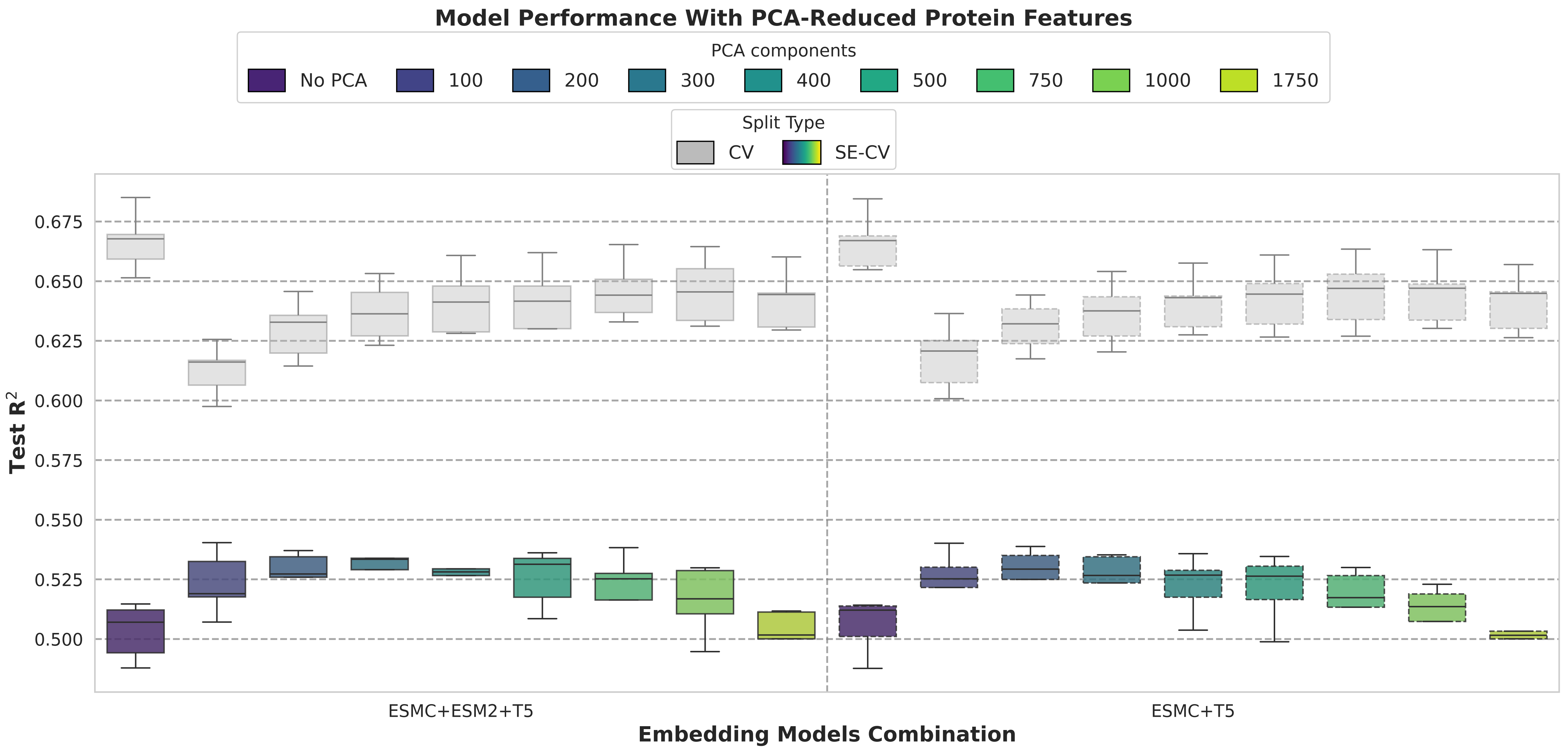}
\par\end{centering}
\centering{}\caption{\label{fig:Effect-of-PCA-based}\textbf{Applying PCA increases $R^{2}$
under SE-CV but decreases $R^{2}$ under standard CV across both protein
representations tested.} $R^{2}$ scores across five folds for Extra
Trees models trained using protein representations with and without
PCA. Results are shown for two protein embedding combinations: ESMC+T5
and ESMC+ESM2+T5. For each setting, PCA was applied with varying numbers
of principal components (100 to 1750), as indicated in the colour
bar. Coloured boxes represent SE-CV results and grey boxes represent
standard CV. }
\end{figure}

As shown in Figure \ref{fig:Effect-of-PCA-based}, using the full
representation (no PCA) yields the highest performance in standard
CV, but performs poorly in SE-CV, indicating overfitting and memorisation
of sequence representations. In contrast, applying PCA improves SE-CV
performance, especially when reducing to 200--400 components (captures
$90$-$95\%$ of test set variance). This trade-off is consistent
across both protein representation settings, suggesting that dimensionality
reduction can improve model robustness by encouraging generalisation
at the cost of in-distribution fit (standard CV performance). As more
components are retained, performance in standard CV increases again,
while SE-CV performance begins to decline. This highlights a trade-off
between retaining detailed information and controlling overfitting.

\section{Discussion}

\subsection{Evaluation Protocols and Generalisation}

\paragraph{Importance of evaluation protocol}

Our results highlight the importance of evaluating model generalisation
to unseen protein sequences, in line with prior findings \cite{kroll_dlkcat_2024kroll_dlkcat_2024}.
The choice of evaluation method strongly influences apparent model
quality. Since SE-CV prevents sequence overlap between folds, it offers
a more realistic evaluation of generalisation and should be adopted
as standard practice for benchmarking kinetic prediction models. 

\paragraph{Overfitting under standard CV}

A key takeaway from our study is that models optimised for generalisation
may appear weaker under random splitting, while models that overfit
the training distribution can appear deceptively strong. For example,
in our PCA experiments (Figure \ref{fig:Effect-of-PCA-based}), not
using PCA led to the highest performance under standard CV but the
worst results under SE-CV. This trend is clearly evident when considering
that under standard CV, KinForm-H and UniKP perform when identical
sequences are present in the training and test sets (\ref{fig:unikp_comp_kcat}).
However, when identical sequences are removed KinForm-L outperforms
across lower similarity bins and matches in the $>90\%$ bin (Figure
\ref{fig:seqsim-Vs-R2}). Furthermore, the per band datapoint counts
in Figure \ref{fig:seqsim-Vs-R2} show that SE-CV yields a more balanced
distribution of test datapoints across similarity bands, whereas standard
CV heavily concentrates them in the high-similarity region.

\subsection{Performance of KinForm Configurations}

We did not include DLKcat in our performance comparisons, as it has
been shown to under-perform significantly relative to UniKP. In the
original UniKP paper, UniKP reports a median $R^{2}$ of $0.68$ across
five random 90/10 train-test splits, compared to $0.5$ for DLKcat.
Moreover, DLKcat exhibits poor generalisation to low-similarity sequences,
with $R^{2}$ dropping to $-0.5$ \cite{kroll_dlkcat_2024kroll_dlkcat_2024}.

We designed KinForm-H and KinForm-L to target complementary objectives:
fit to the training distribution and robust generalisation, respectively.
KinForm-H retains the full 6912-dimensional protein vector and achieves
the strongest performance under standard CV, surpassing UniKP and
matching the final reported performance of the EITLEM model (Figure
\ref{fig:unikp_comp_kcat}). EITLEM is a deep learning framework that
reaches it's peak $R^{2}$ after eight rounds of transfer learning
across $k_{cat}$, $K_{M}$ and $\frac{k_{cat}}{K_{M}}$ datasets.
In contrast, KinForm-H matches this performance (on the same 90/10
split) using a single Extra Trees regressor trained once, with no
pretraining or fine-tuning. This highlights that effective protein
representations can deliver strong performance even when paired with
simple ML models, without requiring complex training pipelines.

By contrast, KinForm-L reduces representation dimensionality via PCA.
Its performance gains are most pronounced in the low-similarity bins,
where it consistently outperforms UniKP. In the $<20
$ identity bin, KinForm-L achieves $44\%$ ($0.09\rightarrow0.13$)
and $100\%$ ($0.075\rightarrow0.15$) higher $R^{2}$ than UniKP
on DLKcat and Shen datasets, respectively (Figure \ref{fig:seqsim-Vs-R2}).
Similar high-margin improvements are observed across similarity bins
$<90\%$. This suggests that KinForm-L is learning transferable biochemical
signals that extend beyond feature memorisation. KinForm-L maintains
UniKP\textquoteright s performance in the $>90\%$ bin, suggesting
that generalisation can be improved without sacrificing in-distribution
accuracy. These results show that applying PCA-based dimensionality
reduction and oversampling of low-similarity sequences (core components
of KinForm-L) can substantially improve generalisation to dissimilar
proteins.

For $K_{\mathrm{M}}$, we evaluate only KinForm-H, as KinForm-L provides
no consistent advantage for this task. Across both standard CV and
SE-CV, KinForm consistently outperforms UniKP by approximately $5\%$
(Figure \ref{fig:km_comp}), indicating that improvements in protein
representation are beneficial for both kinetic parameters. However,
improvements are smaller than those observed for $k_{cat}$ and the
gap between SE-CV and CV results is smaller. We expand on this asymmetry
and its possible causes in Section \ref{subsec:Limitations}.

\subsection{Computational Efficiency }

The ensemble ML models (UniKP and KinForm) train in less than one
minute on CPU and do not require GPUs for fast inference. This makes
them more accessible than deep learning models. They're also more
practical than methods that require protein structure before inference.
Even with a modern GPU, this step can take over 30 seconds per sequence
\cite{marounina_benchmarking_2024marounina_benchmarking_2024}.

\subsection{Methodological Contributions }

\paragraph{Binding-site--weighted pooling}

KinForm departs from prior models in four ways. We replace uniform
mean pooling with binding-site--weighted pooling using residue-level
probabilities from Pseq2Sites. This adjustment emphasises active-site
residues, leading to improved performance in both standard CV and
SE-CV. Concatenating the binding-weighted and global vectors is the
best approach we found (Figure \ref{fig:Effect-of-protein}). While
CV performance improves when using multiple PLLMs and combining pooling
strategies, SE-CV performance does not improve. This suggests that
increasing representation dimensionality may help fit the training
distribution, but it can lead to overfitting and reduced generalisation
to unseen protein sequences.

\paragraph{Intermediate-layer selection}

We implement intermediate-layer selection by systematically evaluating
each encoder layer of the protein language models and selecting the
best-performing one per task. This approach outperforms the default
of using the last layer and reflects the fact that different layers
capture distinct biochemical features (Figure \ref{fig:Layer-wise-performance-of}).
This step offers a principled alternative to default layer choices
and supports more targeted use of pre-trained models.

\paragraph{PCA-based dimensionality reduction}

We apply PCA-based dimensionality reduction to compress high-dimensional
protein embeddings after robust and standard scaling. Reducing from
$3456$ to 200--400 dimensions preserves over $90\%$ of variance
and improves SE-CV performance (Figure \ref{fig:Effect-of-PCA-based}).
This dimensionality reduction improves robustness by limiting the
capacity of models to memorise high-dimensional patterns that may
not generalise beyond the training set. Our results show that while
retaining the full high-dimensional representation leads to the best
performance under standard CV, it performs markedly worse under SE-CV,
indicating substantial overfitting. Applying PCA helps mitigate this
by enforcing a more compact feature space. This trend is consistent
across different protein representation combinations. As the number
of retained components increases beyond this range, standard CV performance
recovers, but SE-CV performance declines. This highlights a clear
trade-off: retaining more components helps preserve task-specific
detail for the training distribution but reintroduces the risk of
overfitting.

\paragraph{Similarity-aware oversampling}

Finally, we introduce similarity-based oversampling, which rebalances
the training data by duplicating sequences from underrepresented similarity
bands. This yields an improvement in the $<20\%$ identity bin and
does not affect performance on high-similarity proteins (Figure \ref{fig:OSvsNO(OS)}).

\subsection{\label{subsec:Limitations}Limitations}

\paragraph{Limited performance on low-similarity sequences}

While KinForm improves generalisation compared to prior models, several
limitations remain, reflecting that representation engineering alone
cannot fully overcome the challenges posed by sparse and biased kinetic
measurements. First, performance on low-similarity sequences is still
limited. In the $<20\%$ identity bin, KinForm-L achieves an $R^{2}$
of approximately $0.20$ on the Shen dataset and $0.15$ on DLKcat.
Although these represent improvements over UniKP, the absolute performance
remains low, indicating that modelling proteins that are dissimilar
to the training set remains a major challenge. 

\paragraph{Substrate-dominant $\mathbf{K_{M}}$ prediction and SMILES limitations}

While the SMILES Transformer was used as the molecular encoder for
all $K_{M}$ predictions, and outperformed alternative methods we
evaluated, the choice of SMILES representation remains a critical
factor for $K_{\mathrm{M}}$. We also observe that the performance
gap between CV and SE-CV is much smaller for $K_{M}$ relative to
$k_{cat}$. Since SE-CV explicitly ensures that no protein sequence
is shared between training and test sets, a smaller performance drop
under this regime suggests reduced reliance on sequence-level generalisation
and increased importance of substrate features. In line with this,
it was shown that models using only substrate features could recover
$\approx90\%$ of the full predictive performance for $K_{M}$ \cite{kroll_deep_2021kroll_deep_2021}.
This work focused on optimising the protein representation; a similar
effort directed toward improving molecular representations, such as
exploring alternatives to global mean pooling over atom embeddings,
may yield further gains especially for $K_{\mathrm{M}}$ prediction. 

\paragraph{Independent prediction of binding-site weights}

The binding-site weights used for pooling are predicted separately
using the Pseq2Sites model. While this improves performance, a limitation
is that the predictor is trained independently. Jointly fine-tuning
a binding-aware representation model or directly incorporating embeddings
from Pseq2Sites into the architecture may yield further improvements

\paragraph{Missing contextual factors}

Our models do not incorporate contextual features such as temperature
or pH, which are known to influence enzyme kinetics \cite{heckmann_machine_2018heckmann_machine_2018a}.
These contextual factors could be integrated with minimal architectural
change, mirroring UniKP-EF. 

\paragraph{Fixed layer selection and potential bias}

Layer-wise results indicate that intermediate transformer layers are
most predictive, suggesting that different layers capture distinct
biochemical features. While we used fixed hyperparameters for consistency,
this may under represent the potential of some layers, particularly
deeper ones, which could benefit from task-specific tuning. Furthermore,
optimal hidden layers were chosen using SE-CV on DLKcat with different
random seeds than other experiments, but since the same dataset was
used some selection bias may remain, and results on the Shen dataset
provide a more unbiased estimate of generalisation.

\subsection{Future Work }

\paragraph{Hyperparameter optimisation and sophisticated pooling }

Additional gains may be achieved through hyperparameter optimisation
of the Extra Trees model. Furthermore, we used binding-site--weighted
pooling as a simple way to improve over global mean pooling. A more
flexible approach could be to use attention maps from pre-trained
models, such as those trained for function prediction or docking,
to weigh residues based on their relevance. This could help capture
more detailed patterns than fixed binding-site scores. 

\paragraph{Synthetic Data}

Synthetic data generation guided by targeted mutations (eg., using
ESM-3 to explore plausible protein variants) may help expand coverage
in the low-similarity subsets. This is especially relevant given the
limited size and diversity of current experimental datasets. Broader,
higher quality data will be critical for training models that generalise
beyond training set proteins and their homologs. 

\paragraph{Richer reaction context}

The data we use is limited to protein--substrate--value pairs, without
additional context. Adding more reaction information, such as cofactors,
full reaction SMILES, temperature, and pH, could help improve prediction.
But this would require higher quality datasets that include these
variables in a consistent and structured way.

\subsection{Conclusions}

We have demonstrated that targeted improvements in protein representation,
including binding-site weighting, intermediate-layer selection, and
dimensionality reduction, enhance the generalisation of kinetics predictors
to low-similarity sequences. These findings suggest that for tasks
like kinetic parameter prediction, where datasets are relatively small
and biological variance is high, investing effort in feature engineering
can be beneficial. More broadly, our results highlight that meaningful
advances require evaluation protocols that reflect out-of-distribution
generalisation.

\section{Code and Data Availability}

All model training code, including the KinForm configurations, experiment
scripts, and tools for training and generating new predictions, is
available at: \url{github.com/Digital-Metabolic-Twin-Centre/KinForm}.
We use the raw data files released by EITLEM \cite{shen_eitlem-kinetics_2024shen_eitlem-kinetics_2024}
and DLKcat \cite{li_deep_2022li_deep_2022a}, both of which source
their kinetic measurements from BRENDA \cite{chang_brenda_2021chang_brenda_2021a}
and SABIO-RK \cite{wittig_sabio-rk_2018wittig_sabio-rk_2018a}.

\section{Acknowledgements}

This work was supported by the European Union's Horizon Europe Framework
Programme {[}grant number 101080997{]}. We also thank members of the
Recon4IMD consortium for their invaluable feedback, in particular
Prof. Brian Marsden, Prof. Wyatt Yue, Dr Mihaela Atanasova and Dr
Thomas McCorvie.\pagebreak{}

\bibliography{KINEFORM_NoInterface.bib}

\addtocontents{toc}{\protect\setcounter{tocdepth}{0}}

\appendix

\part*{Supplementary Materials}

\section{Dataset Analysis}

\subsection{Dataset Statistics}

\begin{figure}[H]
\centering{}\includegraphics[width=1\textwidth]{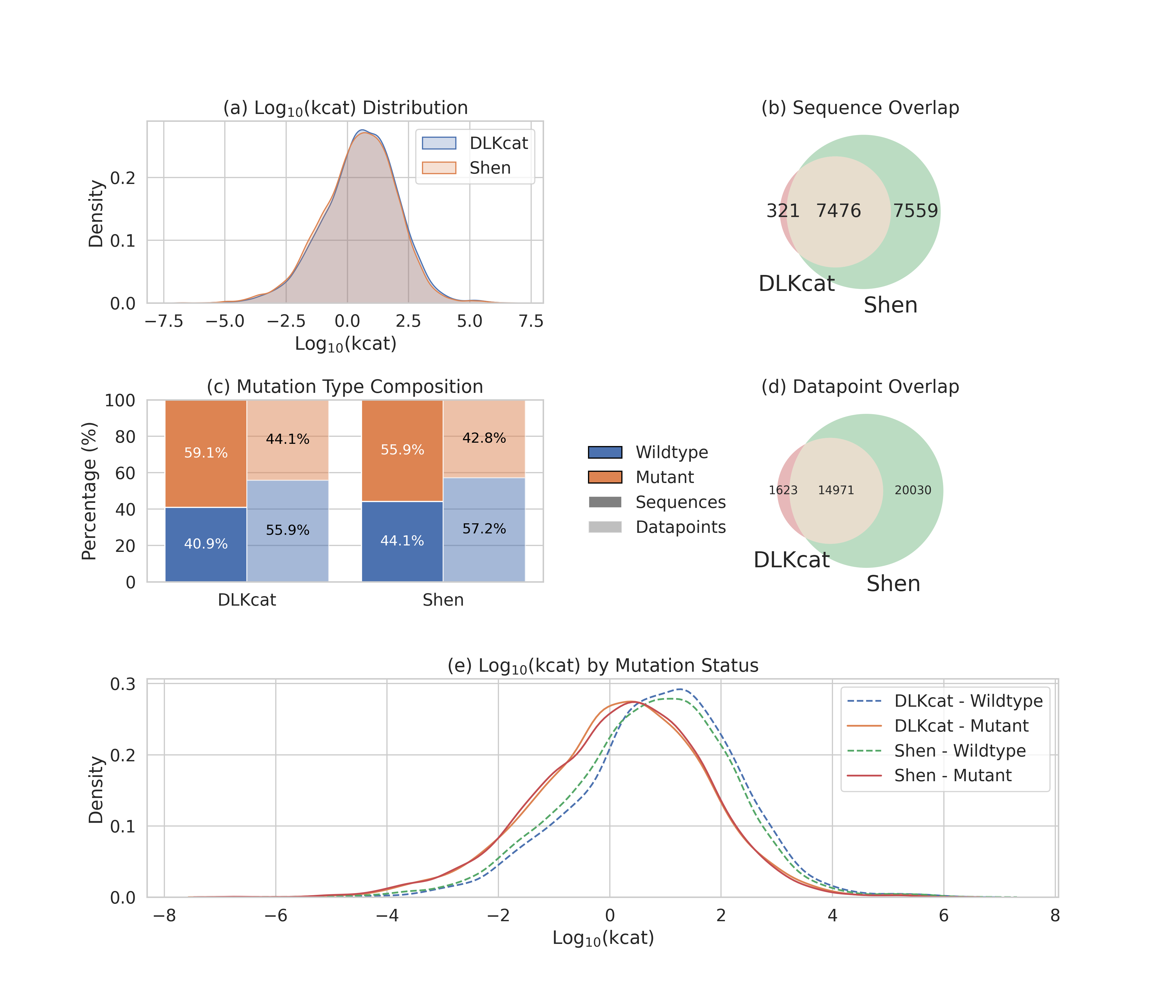}\caption{\textbf{\label{fig:Dataset-comparison}Overall, the Shen dataset is
larger than DLKcat while maintaining similar $\mathbf{k_{cat}}$ distributions
and wild type to mutant ratios, with most DLKcat data contained within
Shen.} \textbf{(a)} Distribution of $\log_{10}k_{cat}$ values in
the DLKcat and Shen datasets. \textbf{(b) }Overlap of unique protein
sequences between DLKcat and Shen. \textbf{(c) }Proportion of wild
type versus mutant sequences and datapoints in each dataset. \textbf{(d)
}Overlap in complete datapoints (defined as identical sequence, SMILES,
and $k_{cat}$ value) between DLKcat and Shen.\textbf{ (e) }Distribution
of $\log_{10}k_{cat}$ values in wild type and mutant subsets of DLKcat
and Shen.}
\end{figure}

\subsection{Sequence Clusters in Each dataset}

To characterise sequence redundancy in the DLKcat and Shen datasets,
protein sequences in each dataset were clustered at $80\%$ identity
and we analysed cluster size distributions. We calculate and plot
diversity metrics seen in the figure below. The Gini coefficient is
defined as $G=1-2\int_{0}^{1}L(F)\,dF$, where $L(F)$ is the Lorenz
curve representing the cumulative fraction of sequences as a function
of the cumulative fraction of clusters $F$. This quantifies inequality
in cluster sizes, where $G=0$ indicates perfect equality (all clusters
contain the same number of sequences), while $G=1$ indicates maximal
inequality (all sequences belong to a single cluster). The Simpson
effective number represents the number of equally sized clusters required
to match the probability that two randomly chosen sequences belong
to the same cluster; and the Shannon effective number gives the number
of equally sized clusters needed to achieve the observed Shannon entropy
of the cluster size distribution.

\begin{figure}[H]
\centering{}\includegraphics[width=1\textwidth]{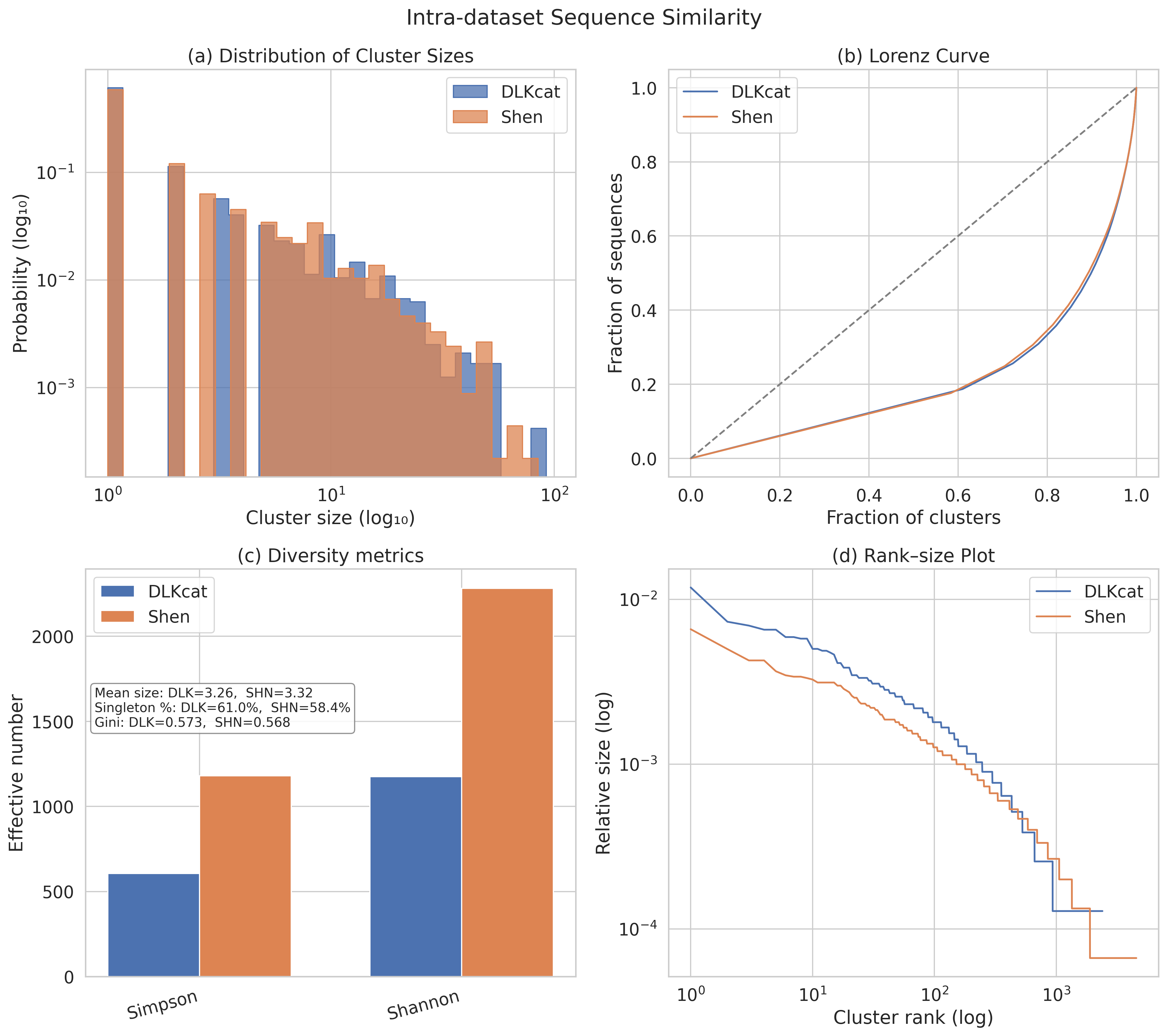}\caption{\textbf{\label{fig:Dataset-sim}Shen and DLKcat datasets exhibit comparable
sequence redundancy patterns, but the Shen dataset spans a broader
sequence space, capturing roughly twice the effective diversity of
DLKcat while maintaining similar levels of redundancy. (a)} The distribution
of cluster sizes on a log--log scale, where the $x$-axis represents
the number of sequences per cluster and the $y$-axis indicates the
probability of observing a cluster of that size. \textbf{(b) }Lorenz
curve, which plots the cumulative fraction of clusters ($x$-axis),
sorted from smallest to largest, against the cumulative fraction of
sequences they contain ($y$-axis). A curve closer to the diagonal
indicates a more even distribution of sequences across clusters, while
curvature away from the diagonal reflects inequality. \textbf{(c)}
Compares diversity metrics across the two datasets. The bar plots
show the Simpson effective number ($\frac{1}{\sum p_{i}^{2}}$) and
the Shannon effective number ($e^{-\sum p_{i}\ln(p_{i})}$), where
$p_{i}$ is the proportion of sequences in cluster $i$. The inset
text reports the mean cluster size, the percentage of clusters containing
only one sequence, and the Gini coefficient of the sequences in each
dataset (DLK and SHN). \textbf{(d) }Presents the rank--size distribution
of clusters on a $\log$ scale. The $x$-axis indicates the rank of
each cluster (with rank 1 being the largest), and the $y$-axis shows
the relative size of the cluster, defined as the number of sequences
it contains as a fraction of all sequences.}
\end{figure}

\section{\label{sec:SMILES-Representation-Results}SMILES Representation Results}

\begin{figure}[H]
\centering{}\includegraphics[width=1\textwidth]{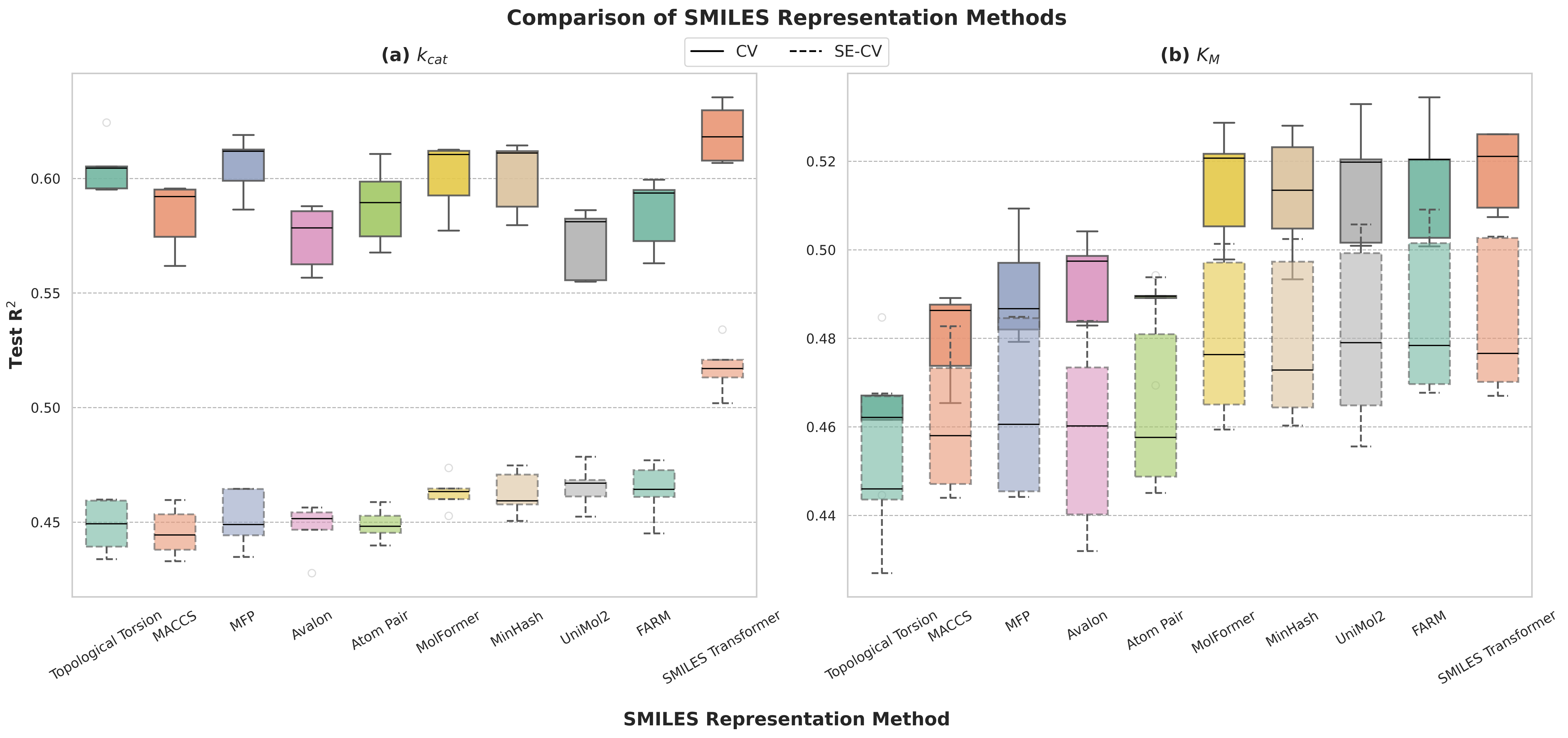}\caption{\label{fig:Comparison-of-SMILES}\textbf{The SMILES Transformer performs
best across both tasks and cross validation methods, with the performance
gap between SE-CV and CV notably smaller for $\mathbf{K_{M}}$ than
for $\mathbf{k_{cat}}$.} $R^{2}$ scores across five folds for different
SMILES representations. Extra Trees is trained with PCA-reduced ESMC+ESM2+T5
(300 components) protein representation. $K_{M}$ experiments are
conducted with ESMC (binding-weighted pooling concatenated with global
pooling) protein representation. Low-opacity, dashed lines represent
SE-CV, while high-opacity, solid lines represent standard CV.}
\end{figure}

\section{PCA Variance}

\begin{figure}[H]
\begin{centering}
\includegraphics[width=0.49\textwidth]{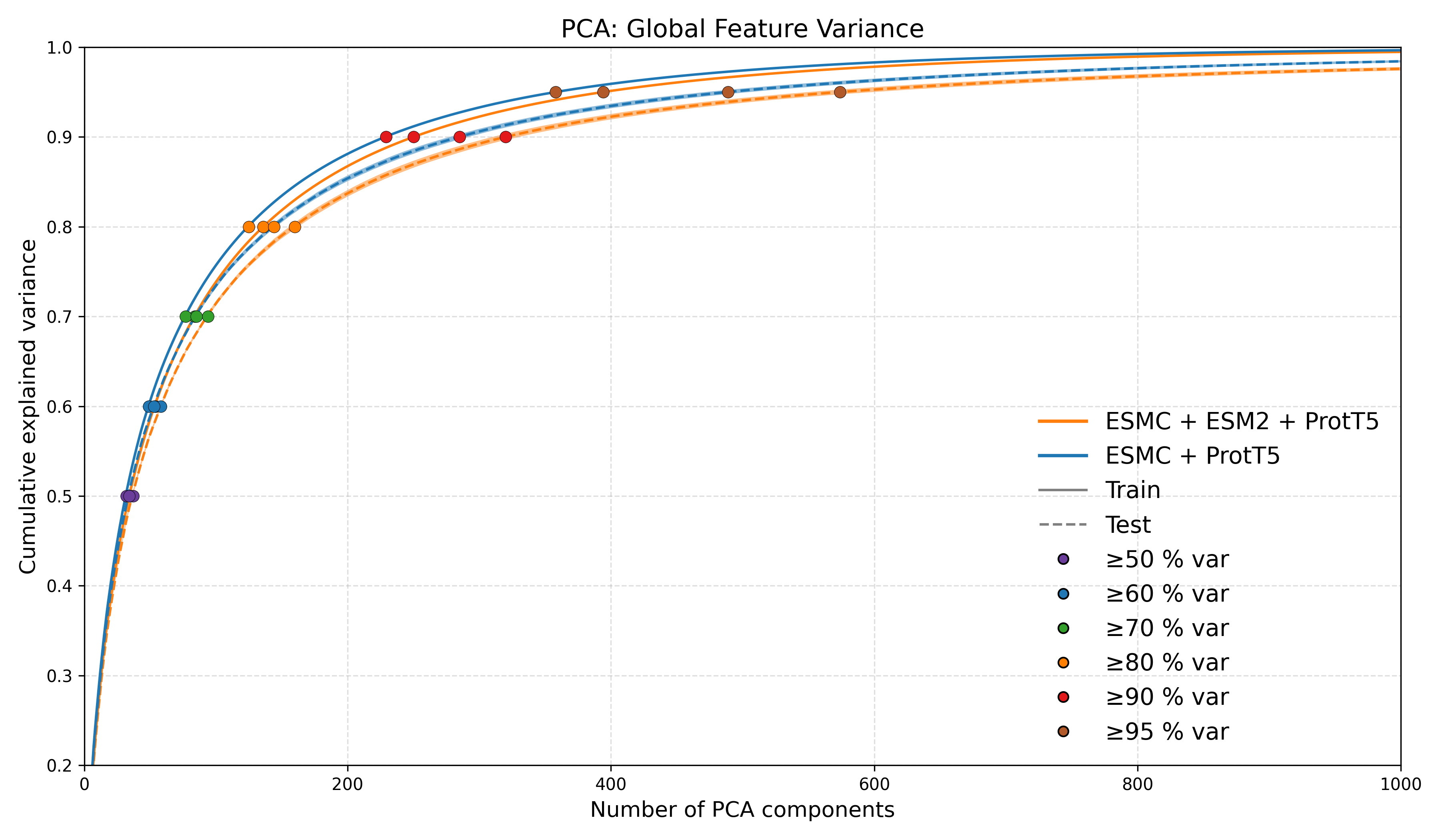}\includegraphics[width=0.49\textwidth]{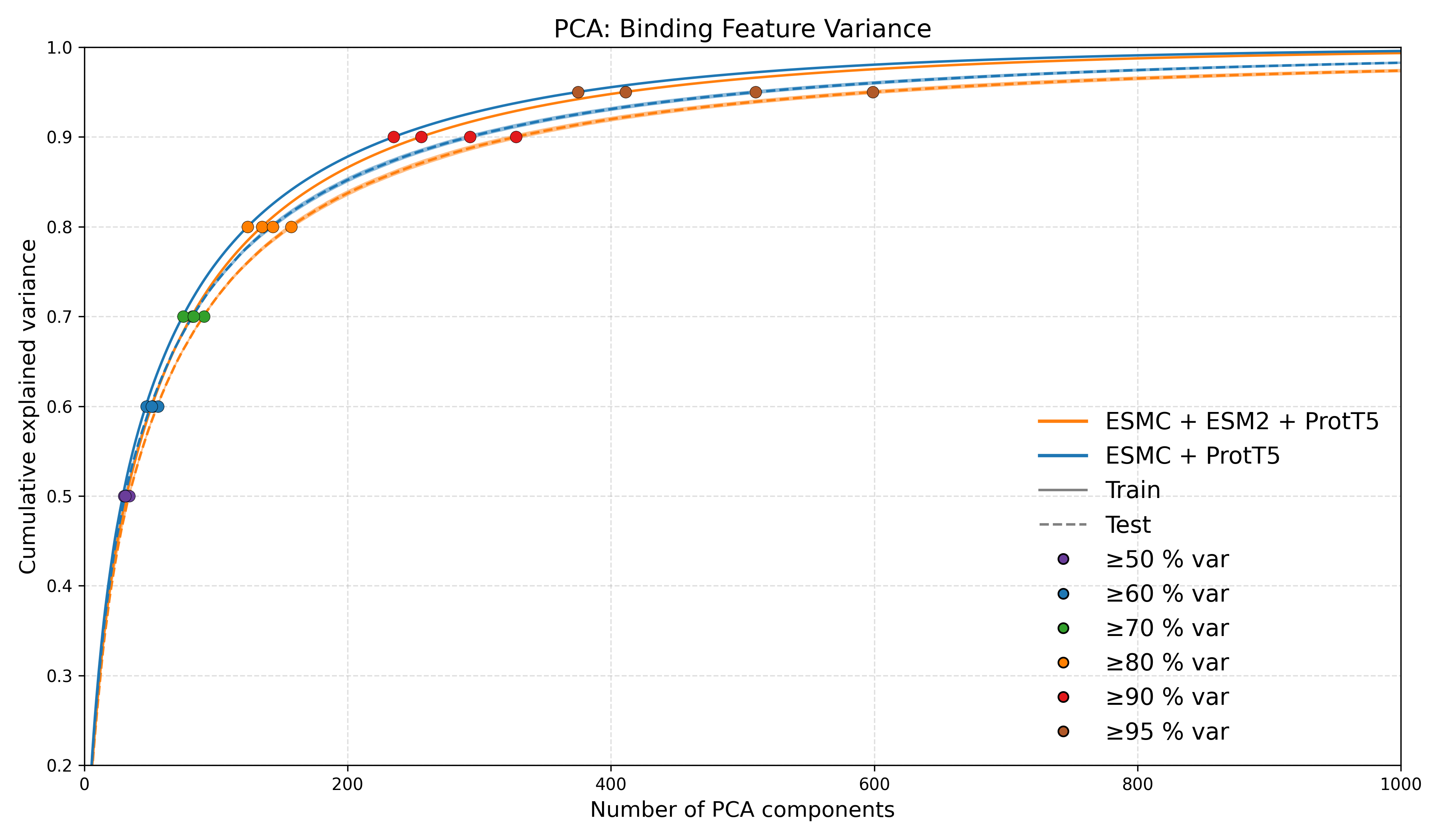}
\par\end{centering}
\centering{}\caption{\label{fig:PCA-Var}\textbf{Explained variance captured by PCA applied
to protein representations.} Each plot shows the cumulative explained
variance as a function of the number of PCA components, computed separately
for the global vector (left) and the binding-weighted vector (right).
PCA is fit on the training set of each SE-CV fold, and mean explained
variance across folds is plotted for both the training set (solid
lines) and test set (dashed lines). Curves are shown for two protein
representation configurations: ESMC+ProtT5 and ESMC+ProtT5+ESM2 (indicated
by colour). Shaded regions show the standard deviation across folds,
though they are small and largely obscured. Coloured dots indicate
the number of components required to reach $50\%$, $60\%$, $70\%$,
$80\%$, $90\%$, and $95\%$ explained variance. Line and dot styles
are explained in the legend. \newline\textbf{ }The original dimension
of the representations is $3456$; capturing $95\%$ of variance on
the test data requires approximately $\sim600$ components for both
global and binding vectors and $\sim400$ components are needed to
capture $95\%$ of the training data variance. }
\end{figure}

\section{\label{sec:SM-PCA-EITLEM-Dataset}PCA Shen Dataset Results}

\begin{figure}[H]
\centering{}\includegraphics[width=0.85\textwidth]{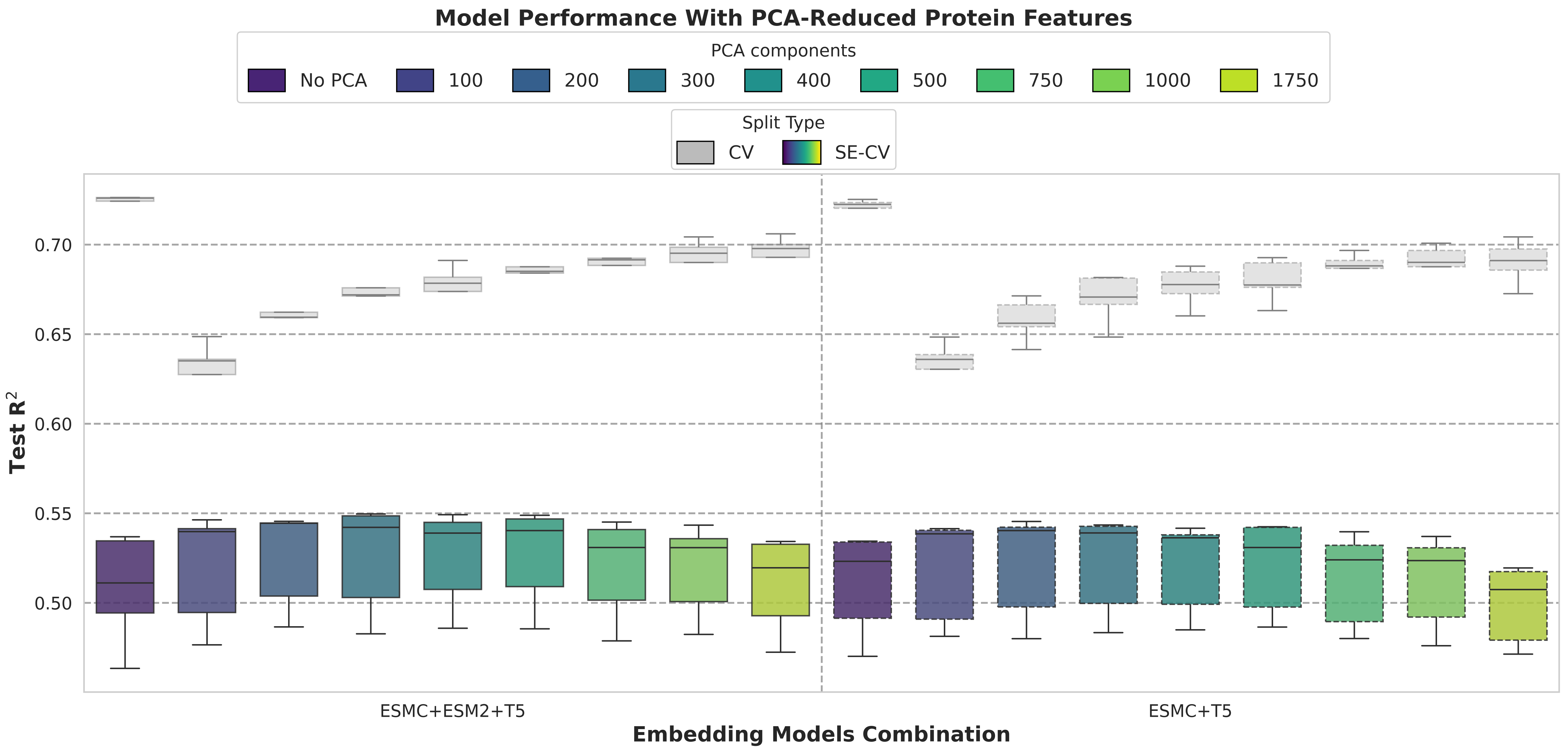}\caption{\label{fig:PCA_GS_EITLEM}\textbf{Effect of PCA-based dimensionality
reduction on predictive performance on the Shen dataset.} This figure
mirrors the analysis shown in Figure \ref{fig:Effect-of-PCA-based}
but applied to the Shen dataset. $R^{2}$ scores across five folds
for Extra Trees models trained with and without PCA using two protein
embedding combinations: ESMC+T5 and ESMC+ESM2+T5. PCA was applied
with varying numbers of components (100 to 1750). Coloured boxes correspond
to SE-CV results; grey boxes correspond to standard CV.}
\end{figure}

\section{\label{sec:prot_rep_gs-dlkcat}Protein Representation DLKcat Dataset
Results}

\begin{figure}[H]
\centering{}\includegraphics[width=1\textwidth]{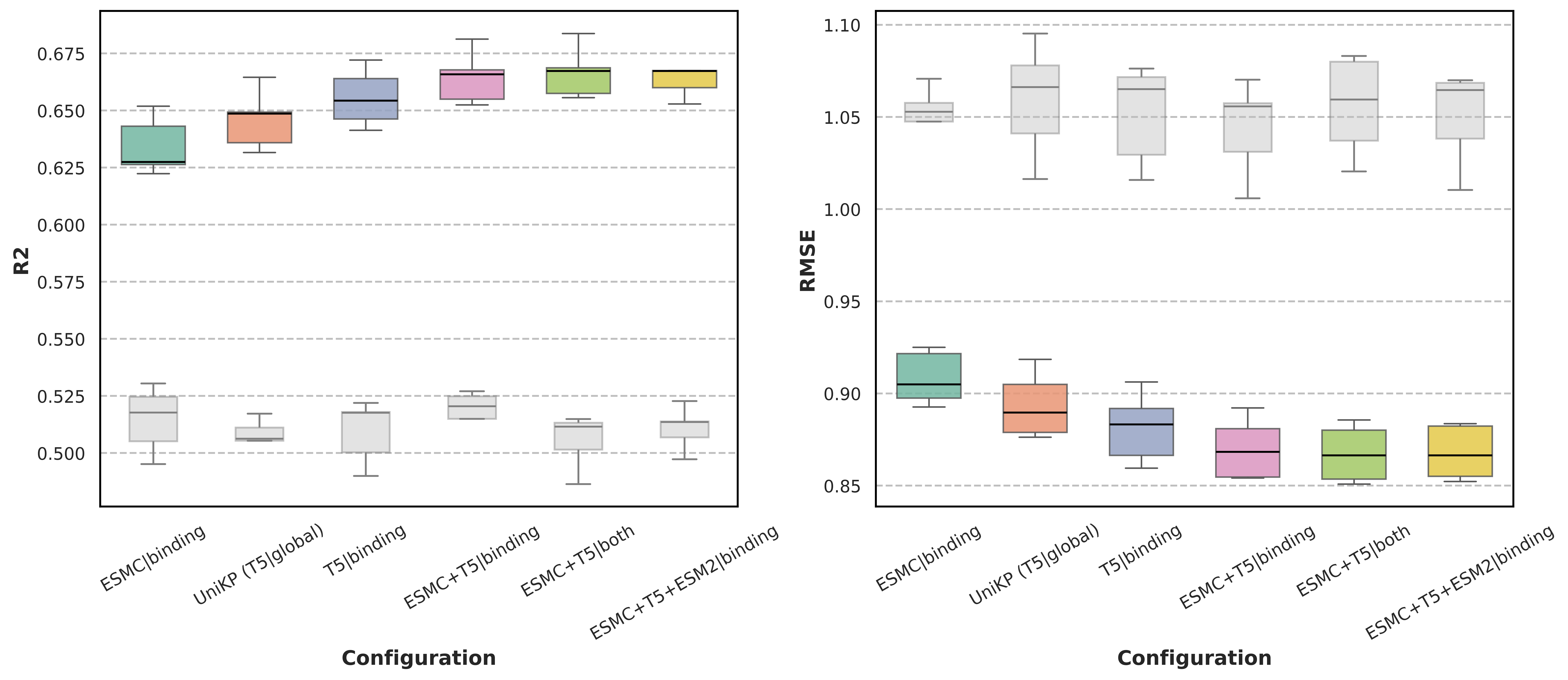}\caption{\label{fig:PCA_GS_EITLEM-1}\textbf{Effect of protein representation
strategies on predictive performance across cross validation settings
on the DLKcat dataset.} This mirrors the analysis shown in Figure \ref{fig:Effect-of-protein}
but applied to the DLKcat dataset. R\texttwosuperior{} scores across
five folds for Extra Trees models trained with different configuration.
Top six configurations are shown. Coloured boxes correspond to standard
CV results; grey boxes correspond to SE-CV.}
\end{figure}

\end{document}